\documentclass[
  10pt,
  aps,
  prb,
  twocolumn,
  preprintnumbers,
  superscriptaddress,
  floatfix
]{revtex4-2}
\usepackage[margin=1in]{geometry}
\usepackage[utf8]{inputenc}
\usepackage{xcolor}
\usepackage{textgreek}
\usepackage[normalem]{ulem}
\usepackage{comment}
\usepackage{pifont}

\usepackage{mathtools}
\usepackage{amssymb}
\usepackage{mathrsfs}
\usepackage{bbold}
\usepackage{bm}
\usepackage{breqn}

\usepackage{graphicx}
\usepackage{adjustbox}
\usepackage{tikz}
\usetikzlibrary{quantikz2}

\usepackage[
  compatibility=false,
  labelfont=bf,
  justification=raggedright,
  singlelinecheck=false
]{caption}
\usepackage{subcaption}

\usepackage{placeins}

\usepackage{booktabs}
\usepackage{makecell}
\usepackage{multirow}
\usepackage{dcolumn}

\begin{document}

\title{Variationally Optimized Imaginary-time Polynomial Filters for Ground State Projection}

\author{Bahman Seifi}
\affiliation{Department of Physics and Physical Oceanography$,$
Memorial University of Newfoundland and Labrador$,$ St. John’s$,$ Newfoundland $\&$ Labrador$,$ Canada A1B 3X7}

\title{Variational optimization of ancilla‑based imaginary time evolution for quantum ground state preparation}
\author{Ibsal Assi}
\affiliation{Department of Physics and Physical Oceanography$,$
Memorial University of Newfoundland and Labrador$,$ St. John’s$,$ Newfoundland $\&$ Labrador$,$ Canada A1B 3X7}

\author{J. P. F. LeBlanc}
\affiliation{Department of Physics and Physical Oceanography$,$
Memorial University of Newfoundland and Labrador$,$ St. John’s$,$ Newfoundland $\&$ Labrador$,$ Canada A1B 3X7}
\affiliation{Compute Everything Technologies Ltd.$,$ St. John's$,$ Newfoundland $\&$ Labrador$,$ Canada}

\date{\today}

\begin{abstract}
In this work, we develop a variational imaginary-time evolution (ITE) framework based on polynomial filtering, derived from an operator-level action principle, which yields an optimized non-unitary projector expressed as a polynomial in the Hamiltonian. Starting from a single-ancilla, first-order imaginary-time update defined by a Taylor expansion and Trotter-Suzuki (TS) decompositions, we show that replacing these approximations with alternative variational formulas substantially improves both accuracy and stability at larger time steps, leading to up to an order-of-magnitude enhancement in the final success probability.  We further derive rigorous error bounds that depend only on static properties of the Hamiltonian, providing practical guidance for selecting the simulation time step. Benchmarks on the transverse-field Ising model demonstrate faster convergence to the ground-state energy and improved robustness compared to standard TS--Taylor ITE, highlighting variational polynomial filtering as a practical route to higher-fidelity ground-state preparation on near-term quantum devices. 
\end{abstract}

\maketitle

\section{Introduction}
Determining the ground state $\ket{\mathrm{GS}}$ of a quantum many-body system is fundamental to predicting its equilibrium properties and phase diagrams~\cite{Fetter2012-jg,Sachdev,resta2020,Wang_PhysRevX2024}. For small system sizes, where the Hamiltonian acts on a manageable Hilbert space, exact diagonalization (ED) provides a numerically exact benchmark~\cite{Sandvik2010-ei,Weinberg2017-fp}. However, as the number of constituents grows, the Hilbert-space dimension expands exponentially, for instance as $2^N$ for a spin-$1/2$ chain of length $N$. With a Hilbert-space dimension $D=2^N$, ED requires $\mathcal{O}(D^2)$ memory and $\mathcal{O}(D^3)$ time, rendering it prohibitive for large $N$. This intrinsic exponential scaling motivates the search for fundamentally different approaches. Quantum computers, which natively operate in exponentially large Hilbert spaces, offer a compelling route to circumvent this bottleneck, spurring the development of dedicated quantum algorithms for efficient ground-state preparation.

Alongside these quantum perspectives, numerous classical strategies have been formulated to overcome ED's exponential bottleneck. Quantum Monte Carlo (QMC) methods achieve scalability via stochastic sampling \cite{Toulouse2016-hc,Lee2022-ii,Huang2002-wt}, yet they suffer from the notorious sign problem when applied to frustrated or fermionic systems \cite{TroyerPRL2005}. Variational wavefunctions and tensor-network states provide flexible ansatz-based approximations \cite{Gould2012-ts,Drake_ch18,Medvidovic2024-yk,whiteDMRG1992,Schollwock2011-wt,Orus2014-ji,SchuchPRL2007,GiovannettiPRL2008,EvenblyPRL2015}; however, their practical performance depends critically on the expressiveness of the ansatz, the underlying entanglement structure, and the achievable bond dimension \cite{Huang1997-ze,RevModPhys.82.277}. These persistent limitations of classical approaches underscore the necessity for quantum-native algorithms, motivating prominent examples such as the variational quantum eigensolver (VQE) and quantum Krylov-subspace (Lanczos) methods \cite{Peruzzo2014-yg,He2024-bv,TangPRXQuantum2021,AnastasiouPhysRevResearch2024,Motta2020-tz}.

Among quantum approaches to GS preparation, imaginary-time evolution (ITE) provides a direct filtering mechanism: applying \(e^{-\tau H}\) to an initial state with nonzero GS overlap exponentially suppresses excited-state components, driving the system toward the GS as the imaginary time \(\tau\) increases \cite{Motta2020-tz,Angles-Castillo2025-rf}. This principle underpins a broad range of applications, from GS certification and energy estimation to studies of low-energy phenomena \cite{Sachdev,Wang_PhysRevX2024,lin2020near,ding2024single,dong2022ground,lin2022heisenberg,belaloui2025ground,wang2025efficient,vojta2003quantum,vojta2006impurity,najarbashi2017quantum}. From a quantum computing perspective, the central obstacle is that \(e^{-\tau H}\) is nonunitary and cannot be implemented as a deterministic gate-based circuit. Existing implementations therefore circumvent this obstruction either through variational approximations that constrain the evolution to a parameterized unitary ansatz \cite{mcardle2019variational,kandala2017hardware,anuar2024operator}, or through probabilistic protocols that embed the nonunitary propagator into larger unitary constructions using Trotter decompositions, linear combinations of unitaries (LCU), or polynomial expansions \cite{leadbeater2024non,yi2025probabilistic,zhang2025quantum}. In the latter category, Yi \textit{et al.}~\cite{yi2025probabilistic} recently introduced a circuit-level probabilistic method tailored to Pauli-product Hamiltonians, based on truncated Taylor expansions and postselected LCU circuits. Despite these advances, practical efficiency remains critically constrained by the trade-off between approximation accuracy and success probability: reducing the expansion error demands finer time steps or higher expansion orders, which inevitably increases circuit depth, measurement overhead, and critically lowers the post-selection success rate. Balancing these competing factors therefore represents a central design challenge for ITE-based quantum algorithms.

To address this inherent trade-off between approximation accuracy and post-selection success probability, we introduce an action-based variational approach (ABVA) for ancilla-assisted probabilistic ITE. Rather than relying on fixed coefficients in standard Taylor or Trotter-Suzuki (TS) decompositions, our approach variationally optimizes these parameters while preserving the original circuit architecture. We begin with a baseline construction based on a first-order Taylor update, $1-\tau H$, combined with a second-order TS decomposition for the real-time block. Our ansatz generalizes this structure by promoting the fixed coefficients to variational parameters, replacing the first-order Taylor form $1-\tau H$ with the generalized action-based variational polynomial ansatz (ABVPA), $c_0(\tau) + c_1(\tau) H$, and analogously replacing the conventional second-order TS decomposition $e^{-itA/2}e^{-itB}e^{-itA/2}$ with the action-based variational TS ansatz (ABVTSA), $e^{id_0(t)A}e^{id_1(t)B}e^{id_2(t)A}$. The coefficients $c_0(\tau)$, $c_1(\tau)$, and $d_j(t)$ are determined through the action-based variational principle \cite{Vogl2025-ku,Assi2026}. Critically, this framework retains the same gate sequence as conventional implementations, yet shifts the burden from static analytical approximations to parameter optimization. As a direct consequence, it alleviates the accuracy-probability bottleneck: the variational parameters deliver faster convergence with higher final fidelity while simultaneously improving the post-selection success probability, substantially reducing the number of circuit shots required to achieve a given statistical precision.

For clarity, we use ABVA as an umbrella term for the overall variational framework developed in this work. The variational polynomial construction is denoted ABVPA, whereas the variational TS construction is denoted ABVTSA. Accordingly, ABVA refers to the complete ancilla-assisted ITE framework when both ABVPA and ABVTSA are employed.

The remainder of this paper is organized as follows. Section~\ref{sec:methodology} details our methodology, covering both the theoretical framework and its quantum-circuit implementation. Section~\ref{sec:error} provides an analysis of the error bounds and resource requirements of the proposed constructions. In Section~\ref{sec:result}, we benchmark our approach on the one-dimensional Ising model with longitudinal and transverse fields. Finally, Section~\ref{sec:conclusions} concludes with a summary and an outlook on future directions.

\section{Methodology}
\label{sec:methodology}
\subsection{Imaginary-Time Evolution and Ancilla-Based LCU Implementation}
\label{sec:ITE-method}
ITE provides a systematic approach to project a quantum state onto the GS of a Hamiltonian $H$. For a time-independent $H$, the imaginary-time evolution operator $U(\tau)$ satisfies
\begin{equation}
\label{eq:SE}
\frac{dU}{d\tau} + H U = 0,
\end{equation}
with solution $U(\tau) = e^{-\tau H}$. Since $H$ is Hermitian, its eigenstates $\{\ket{E_j}\}$ (with \(\ket{E_0}\equiv\ket{\mathrm{GS}}\)) form a complete basis of the Hilbert space.

Starting from an initial state $\ket{\psi(0)} = \sum_j \omega_j \ket{E_j}$ with non-zero GS overlap ($\omega_0 \neq 0$), the evolved state is
\begin{equation}
\ket{\psi(\tau)} \propto e^{-\tau H}\ket{\psi(0)} = \sum_j \omega_j e^{-E_j \tau}\ket{E_j}.
\end{equation}
As $\tau \to \infty$, excited-state contributions are exponentially suppressed, yielding $\ket{\psi(\infty)} \propto \ket{E_0}$. This filtering mechanism underlies the effectiveness of ITE.

Although the formulation developed below applies to a general Hamiltonian $H$, we use the quantum Ising model (QIM) as a representative benchmark:
\begin{equation}
\label{eqn:QIM}
H = J\sum_{j=1}^{N-1}S_j^zS_{j+1}^z + h_x\sum_{j=1}^{N}S_j^x + h_z\sum_{j=1}^{N}S_j^z,
\end{equation}
where $J$ is the nearest-neighbor Ising coupling, while $h_x$ and $h_z$ denote the transverse and longitudinal fields, respectively. We define $S_j^\alpha=\sigma_j^\alpha/2$, with $\alpha\in{x,y,z}$, where $\sigma_j^\alpha$ is the corresponding Pauli operator acting on site $j$. Unless stated otherwise, we use the benchmark QIM with open boundary conditions, and $J=h_x=h_z=1$ throughout this work.

Direct implementation of $e^{-\tau H}$ is challenging from both classical- and quantum-simulation perspectives. In classical simulations, explicitly constructing and exponentiating the Hamiltonian becomes computationally expensive for many-body systems because the Hilbert-space dimension grows exponentially with system size; for an $N$-qubit system, the dimension is $2^N$ \cite{Motta2020-tz}. On gate-based quantum hardware, the difficulty is instead that $e^{-\tau H}$ is nonunitary and therefore cannot be implemented directly as a deterministic unitary gate \cite{terashima2005nonunitary,liu2021probabilistic}. In the ancilla-assisted approach developed below, the $N$-qubit system register is coupled to an ancilla qubit, which is prepared in an appropriate superposition and subsequently measured. Postselection on the desired measurement outcome induces an effective nonunitary transformation on the system \cite{kosugi2022imaginary,leadbeater2024non}. This provides a probabilistic gate-based implementation of discrete ITE, with the short-time approximation and controlled-unitary construction derived below.

\begin{widetext}
\begin{center}
\includegraphics[width=\textwidth]{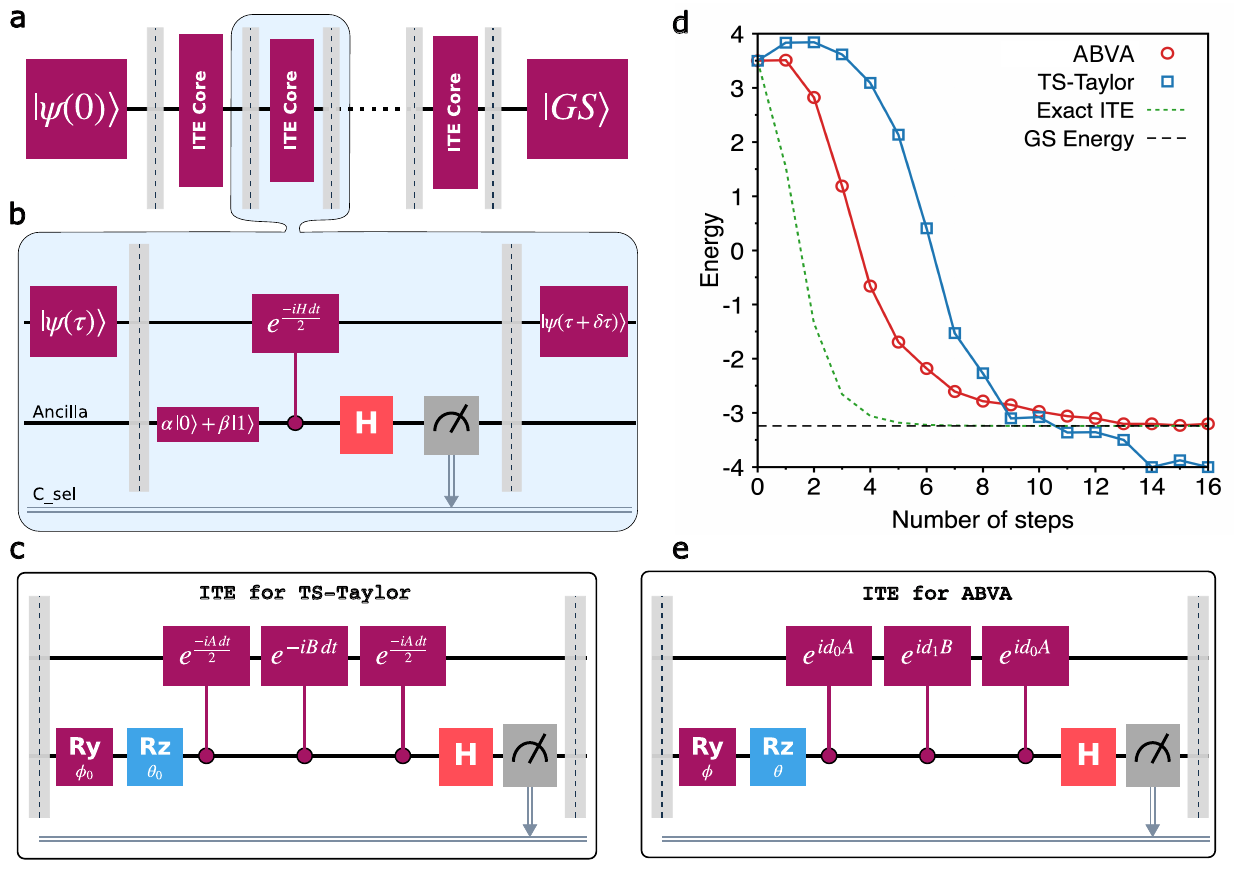}
\captionof{figure}{Ancilla-assisted implementation of ITE. (a) Repeated application of the ITE core filters an initial state $\ket{\psi(0)}$ toward the GS $\ket{GS}$. (b) Measurement-induced LCU circuit for a single ITE step. An ancilla prepared as $\alpha\ket{0}+\beta\ket{1}$ controls the real-time evolution block. After the Hadamard gate, measurement, and post-selection, the system is updated from $\ket{\psi(\tau)}$ to $\ket{\psi(\tau+\delta\tau)}$. (c) Taylor-based ITE circuit in which the controlled real-time propagator is implemented using a second-order Trotter--Suzuki decomposition. (d) Energy evolution obtained using the TS--Taylor and optimized implementations, compared with exact ITE and the exact GS energy for the benchmark Hamiltonian in Eq.~\eqref{eqn:QIM} with $N=5$. The TS--Taylor results are discussed in the present section, while the optimized results are discussed in the following section.(e) Optimized ITE circuit in which the ancilla-preparation and unitary-block parameters are determined using the action-based formulation.}
\label{fig:ITE_circuit}
\end{center}
\end{widetext}
 The ancilla-assisted implementation of ITE considered here is summarized in Fig.~\ref{fig:ITE_circuit}. As illustrated in Fig.~\ref{fig:ITE_circuit}(a), the ITE core is applied repeatedly to project the initial state toward the GS. We develop this procedure in four steps.  We first derive the target linear combination that approximates a short imaginary-time update. We then show how this update is implemented probabilistically using the ancilla-assisted LCU circuit in Fig.~\ref{fig:ITE_circuit}(b), which combines controlled real-time evolution with measurement and post-selection. Next, we construct the real-time propagator using the TS and Taylor (TS--Taylor) decomposition shown in Fig.~\ref{fig:ITE_circuit}(c). Finally, we derive the single-step and cumulative success probabilities. The resulting TS--Taylor energy evolution for QIM~ \ref{eqn:QIM}, is illustrated in Fig.~\ref{fig:ITE_circuit}(d). The optimized construction in Fig.~\ref{fig:ITE_circuit}(e), together with its corresponding curve in panel (d), is presented in the next subsection.

\subsubsection{Single-step update and target linear combination}
For a sufficiently small imaginary-time interval $\delta\tau$, the propagator can be approximated as

\begin{equation}
e^{-H\delta\tau}\approx c'_0+c'_1H.
\end{equation}
where $c'_0=1$ and $c'_1=-\delta\tau$ for the first-order Taylor expansion. Provided that the initial state has a nonzero GS overlap, successive applications of this short-time approximation suppress the excited-state components and drive the state toward the GS without explicitly constructing the dense matrix exponential, as illustrated in Fig.~\ref{fig:ITE_circuit}(a).

More generally, a discretized ITE update can be written to first order as
\begin{equation}
\label{eq:ITE_discrete}
\ket{\psi(\tau+\delta\tau)}\approx\frac{1}{\mathcal{N}}(c'_0 I+ c'_1 H )\ket{\psi(\tau)}.
\end{equation}
where $\mathcal{N}$ is the normalization factor. Meanwhile, a short real-time evolution operator obeys
\begin{equation}
\label{eq:TE_discrete}
e^{-iH\delta t}\approx (c_0 I +c_1H),
\end{equation}
with $c_0=1$ and $c_1=-i\delta t$ for Taylor expansion. Using Eq.~(\ref{eq:TE_discrete}) we obtain the short-time identity
\begin{equation}
H\approx \frac{1}{c_1} e^{-iH\delta t}-\frac{c_0}{c_1}I.
\end{equation}
Identifying $\delta\tau=\delta t$ and substitusing this expression into Eq.~\eqref{eq:ITE_discrete} up to a global phase, yields the update

\begin{equation}
\label{eq:target_map}
\ket{\psi(\tau+\delta\tau)} \approx \frac{1}{\mathcal{N}'}(I + r e^{-iH\delta \tau})\ket{\psi(\tau)}, 
\end{equation}
where $\mathcal{N}'=\mathcal{N}(\frac{c_1}{c'_0c_1-c_0c'_1})$ and $r$ is given by
\begin{equation}
\label{eqe:ancr}
r = \frac{c_1'}{c_0'c_1 - c_1'c_0}.
\end{equation}
For Taylor expansion, this simplifies to
\begin{equation}
r= -\frac{1+i}{2}.
\end{equation}

Equation~\ref{eq:target_map} expresses the short imaginary-time update as a LCU, namely $I$ and $U_{\mathrm{R}}=e^{-iH\delta t}$. This decomposition can therefore be implemented using the measurement-induced circuit shown in Fig.~\ref{fig:ITE_circuit}(b).

\subsubsection{Ancilla-based ITE on a quantum computer}
We now derive the post-selected circuit for a single ITE step that realizes the measurement-induced LCU map, as shown in Fig. \ref{fig:ITE_circuit}(b). The controlled real-time evolution block is introduced explicitly within this construction.

The initial state includes the system state $\ket{\psi(\tau)}$ and ancilla state $\ket{\chi}$ is given by
\begin{equation}
\ket{\phi_{in}}=\ket{\chi}\otimes\ket{\psi(\tau)},
\end{equation}
where the ancilla qubit is parametrized as
\begin{equation}
\label{eqn:ancilla-qubit}
\ket{\chi} = \alpha\ket{0} + \beta\ket{1},
\,\,
|\alpha|^2 + |\beta|^2 = 1,
\end{equation}
this ancilla qubit is prepared using the rotations $R_y(\phi)$, with $\phi=2\arctan(|r|)$, and $R_z(\theta)$, with $\theta=\arg(r)$. For the first-order Taylor approximation, these angles are

\begin{equation}
\begin{aligned}
& \phi_0=2\arctan(\frac{1}{\sqrt{2}}),\\
& \theta_0=\frac{5\pi}{4}.
\end{aligned}  
\end{equation}
The schematic representation of these rotation gates applied to the ancilla qubit is shown in Fig~\ref{fig:ITE_circuit}(c))

Let $U(\delta t)\approx e^{-iH\delta t}$ denote a short real-time evolution block (implemented, e.g., via a Trotter--Suzuki step).  
A controlled-$U$ gate is defined as
\begin{equation}
\Lambda(U)=\ket{0}\bra{0}\otimes I +\ket{1}\bra{1}\otimes U.
\end{equation}
After applying $\Lambda(U)$, the resulting state will be
\begin{equation}
\Lambda(U)\ket{\phi_{in}}=\alpha \ket{0} \ket{\psi(\tau)}+\beta \ket{1} (U\ket{\psi(\tau)}),
\end{equation}
to obtain the desired ITE update, we measure the ancilla in the $X$ basis and post-select the $\ket{+}$ outcome, where
$\ket{+}=(\ket{0}+\ket{1})/\sqrt{2}$. This is implemented by applying a Hadamard gate (H)
\begin{equation}
    {\rm H}=\frac{1}{\sqrt{2}}
\begin{pmatrix}
1 & 1\\
1 & -1
\end{pmatrix}
\end{equation}
to the ancilla before measurement, which gives
\begin{equation}
\begin{aligned}
(\text{H}\otimes I)\Lambda(U)\ket{\phi_{in}} &= \frac{1}{\sqrt{2}}
\ket{0}(\alpha I+\beta U )\ket{\psi(\tau)} \\
&\quad +\frac{1}{\sqrt{2}}
\ket{1}(\alpha I-\beta U )\ket{\psi(\tau)}.
\end{aligned}
\end{equation}
Post-selecting on the ancilla outcome $\ket{0}$ yields the (unnormalized) system state
\begin{equation}
\ket{\phi_{out}} = \frac{1}{\sqrt{2}}(\alpha I + \beta U)\ket{\psi(\tau)}.
\end{equation}

Choosing $\alpha=1/\sqrt{1+|r|^2}$ and $\beta=r/\sqrt{1+|r|^2}$, this becomes
\begin{equation}
\label{eq:Kraus}
\ket{\phi_{out}}=\mathcal{K}_r\ket{\psi(\tau)},
\qquad
\mathcal{K}_r=\frac{I+rU}{\sqrt{2(1+|r|^2)}},
\end{equation}
here $\mathcal{K}_r$ is the Kraus operator associated with the post-selected ancilla outcome $\ket{0}$ \cite{kraus1987complementary}. The normalised output state $\ket{\phi_{out}}$ is
\begin{equation}
\ket{\phi_{out}}=\frac{\mathcal{K}_r\ket{\psi(\tau)}}{\sqrt{\bra{\psi(\tau)}\mathcal{K}_r^\dagger \mathcal{K}_r\ket{\psi(\tau)}}},
\end{equation}
and, comparing with Eq.~\eqref{eq:target_map}, we find
\begin{equation}
\ket{\phi_{out}}=\ket{\psi(\tau+\delta\tau)},
\end{equation}
up to a global phase. This completes a single ITE step.

\subsubsection{Trotter--Suzuki block for $U(\delta\tau)$}
To implement the real-time propagator $U_{\mathrm{R}}(\delta t)=e^{-iH\delta t}$ appearing in the LCU map, we use the second-order TS decomposition shown in Fig.~\ref{fig:ITE_circuit}(c). For a general nearest-neighbor Hamiltonian,
\begin{equation}
H = \sum_{\ell} H_\ell,
\end{equation}
where each $H_\ell$ may comprise a set of mutually commuting terms, simplifying the corresponding circuit implementation. The multi-term extension of the second-order symmetric Strang splitting is \cite{strang1968construction}

\begin{equation}
U(\delta t)\approx
\Bigg(\prod_{\ell=1}^{L-1} e^{\frac{-iH_\ell\delta t}{2}}\Bigg)
e^{-i\delta t H_L}
\Bigg(\prod_{\ell=L-1}^{1} e^{\frac{-iH_\ell\delta t}{2}}\Bigg).
\end{equation}

For the QIM~(\ref{eqn:QIM}), we choose the partition $H=A+B$, with 
\begin{equation}
\label{eqn:HamiltonianParts}
A=J\sum_{j=1}^{N-1}S_j^zS_{j+1}^z
+h_z\sum_{j=1}^N S_j^z,\ \ \ \ 
B=h_x\sum_{j=1}^N S_j^x ,
\end{equation}
such that the terms within $A$ and within $B$ commute. The corresponding second-order TS form is
\begin{equation}
\label{eqn:TS_formula}
U(\delta t)=e^{i d_0(\delta t) A}e^{i d_1(\delta t) B}e^{i d_2(\delta t) A},
\end{equation}
where $d_0=d_2=-\delta t/2$ and $d_1=-\delta t$. Each exponential can be implemented using single or two-qubit rotations:
\begin{equation}
\begin{aligned}
e^{-i\theta S_j^x} &= R_X(\theta),\\
e^{-i\theta S_j^z} &= R_Z(\theta),\\
e^{-i\theta S_j^zS_{j+1}^z} &= R_{ZZ}(\theta/2).
\end{aligned}
\end{equation}
These rotations provide a hardware-efficient implementation of $U(\delta t)$. Together with the ancilla preparation, measurement,
and post-selection, they form the complete TS--Taylor ITE circuit shown in Fig.~\ref{fig:ITE_circuit}(c).

Figure~\ref{fig:ITE_circuit}(d) illustrates GS projection by ITE as a function of the number of imaginary-time steps for the benchmark
QIM in Eq.~\eqref{eqn:QIM}, with $N=5$ and a fixed step size of $\delta\tau=0.4$. Provided that the initial state has a nonzero GS overlap, successive ITE updates suppress the excited-state components and drive the evolved state toward the GS. Accordingly, the exact ITE energy, shown by the green dotted curve, converges to the exact GS energy indicated by the horizontal dashed line.

\subsubsection{Per-step and total success probabilities}
Application of the Kraus operator \eqref{eq:Kraus} to $\ket{\psi(\tau)}$ yields a per-step success probability \cite{aharonov2009multiple}
\begin{equation}
\label{eqn:Err1}
p_{\mathrm{step}}
=\|\mathcal{K}_r \ket{\psi(\tau)}\|^2
=\frac{1}{2} 
+ \frac{\mathrm{Re}\!\left[r\langle U \rangle_\psi\right]}{1+|r|^2},
\end{equation}
where $\langle U\rangle_\psi=\bra{\psi(\tau)}U\ket{\psi(\tau)}$. For the special case $r = -(1+i)/2$ , in the limit of small $\delta\tau$ we obtain 
\begin{equation}
\label{eqn:Err2}
p_{\mathrm{step}}
= \frac{1}{6}
- \frac{1}{3}\langle H\rangle_\psi\,\delta\tau
+ \frac{1}{6}\langle H^2\rangle_\psi\,\delta\tau^2
+ \mathcal{O}(\delta\tau^3),
\end{equation}
which shows the per-step success in the limit of $\delta\tau\rightarrow 0$ is 
$p_{\mathrm{step}}=1/6$.

For $n=\tau/\delta\tau$ imaginary-time steps, the total success probability is
\begin{equation}
P_{\mathrm{succ}} 
= \prod_{\ell=1}^{n} p_{\mathrm{step}}^{(\ell)} .
\end{equation}
Since each step is post-selected, $P_{\mathrm{succ}}$ decreases exponentially with the number of steps unless the initial state already has a sufficiently large GS component. To see this, we expand the initial state in the eigenbasis of the Hamiltonian,
($\ket{\psi(0)}=\sum_j \omega_j \ket{E_j}$)
The initial overlap with the GS is therefore
\begin{equation}
|\braket{E_0|\psi(0)}|^2=|\omega_0|^2 .
\end{equation}
If $|\omega_0|^2$ is very small, many imaginary-time steps are required before the GS contribution dominates, which reduces the total post-selection probability. This motivates preparing an initial state with a reasonably large overlap with $\ket{E_0}$.

In practice, the time step $\delta\tau$ should also be chosen moderately. Very small $\delta\tau$ gives only a weak imaginary-time update per successful step, so many post-selected steps are required and $P_{\mathrm{succ}}$ becomes exponentially suppressed. On the other hand, very large $\delta\tau$ reduces the accuracy of the short-time approximation to the imaginary-time propagator and can increase TS and Taylor-expansion errors. Thus, a moderate time step is chosen as a compromise between convergence speed, circuit accuracy, and post-selection success.

In the next section, we introduce the action-based variational approach (ABVA) for the imaginary-time evolution operator and review variational approximations for the real-time propagator. The ABVA yields renormalized coefficients for the Taylor and TS expansions, achieving higher accuracy while retaining the original circuit depth.

\subsection{Alternative variational approximations}
\label{sec:var_ITE}
In this section, we provide ABVA for ITE and the real-time counter parts. To begin, we introduce the following action
\begin{equation}
    S=\int d\tau\,\mathrm{Tr}\left[U^\dagger (\dot{U}+HU)\right].
\end{equation}
By extremizing $S$ with respect to the entries of $U^\dagger$ (or $U$), result in the Schr\"odinger equation \eqref{eq:SE}. For a given ITE ansatz $U_a(c_j'(\tau))$ parametrized by parameters $\{c_j'(\tau)\}$. Extremizing the action $S$ for such ansatz results in a set of equations of motion (EOM) for the variational parameters,
\begin{equation}
\label{eqn:diff_eq_var_parms}
\sum_{k}g_{jk}\dot{c}_k' +F_j = 0,
\end{equation}
where $g_{jk} = \mathrm{Tr} \left[ \left( \frac{\partial U_{a}}{\partial c_j'} \right)^\dagger \frac{\partial U_{a}}{\partial c_k'} \right]$ 
is the quantum geometric tensor (QGT), and 
$F_j = \mathrm{Tr} \left[ \left( \frac{\partial U_{a}}{\partial c_j'} \right)^\dagger H U_a \right]$ 
is the generalized force \cite{Assi2026,Vogl2025-ku}. 

Using the ABVPA, $U_a(c_j'(\tau))=\sum_jc_j'(\tau)H^j$, the EOM become
\begin{equation}
\label{eqn:EOM_ITE_poly_ansatz}
    \sum_{k}\mu_{j+k}\dot{c}_k' +\sum_k\mu_{j+k+1}c_k' = 0,
\end{equation}
where $\mu_n=\mathrm{Tr}[H^n]$. Thus, we can now replace the linear Taylor expansion $e^{-\tau H}\approx 1-\tau H$ with linear ABVPA, $e^{-\tau H}\approx c_0'(\tau)+c_1'(\tau)H$. Solving Eq. \eqref{eqn:EOM_ITE_poly_ansatz} for this linear ansatz gives
\begin{equation}
\label{eqn:cs_imaginaryTE}
\begin{aligned}
c_0'(\tau)& = e^{-\frac{\mu_3}{2\mu_2} \tau} \left[ \cosh(\Omega \tau) + \frac{\mu_3}{2\mu_2\Omega} \sinh(\Omega \tau) \right],\\
c_1'(\tau) &= -\, e^{-\frac{\mu_3}{2\mu_2}\tau} \frac{\sinh(\Omega \tau)}{\Omega},
\end{aligned}
\end{equation}
where $\Omega = \frac{1}{2}\sqrt{ \left(\frac{\mu_3}{\mu_2}\right)^2 + 4\frac{\mu_2}{\mu_0} }$. For a given Hamiltonian one only needs to compute the moments $\mu_n$. As an illustration, we derived those moments for our quantum Ising model \eqref{eqn:QIM} as shown in Appendix \ref{app:Traces}. 

\begin{figure}
    \centering
    \includegraphics[width=\linewidth]{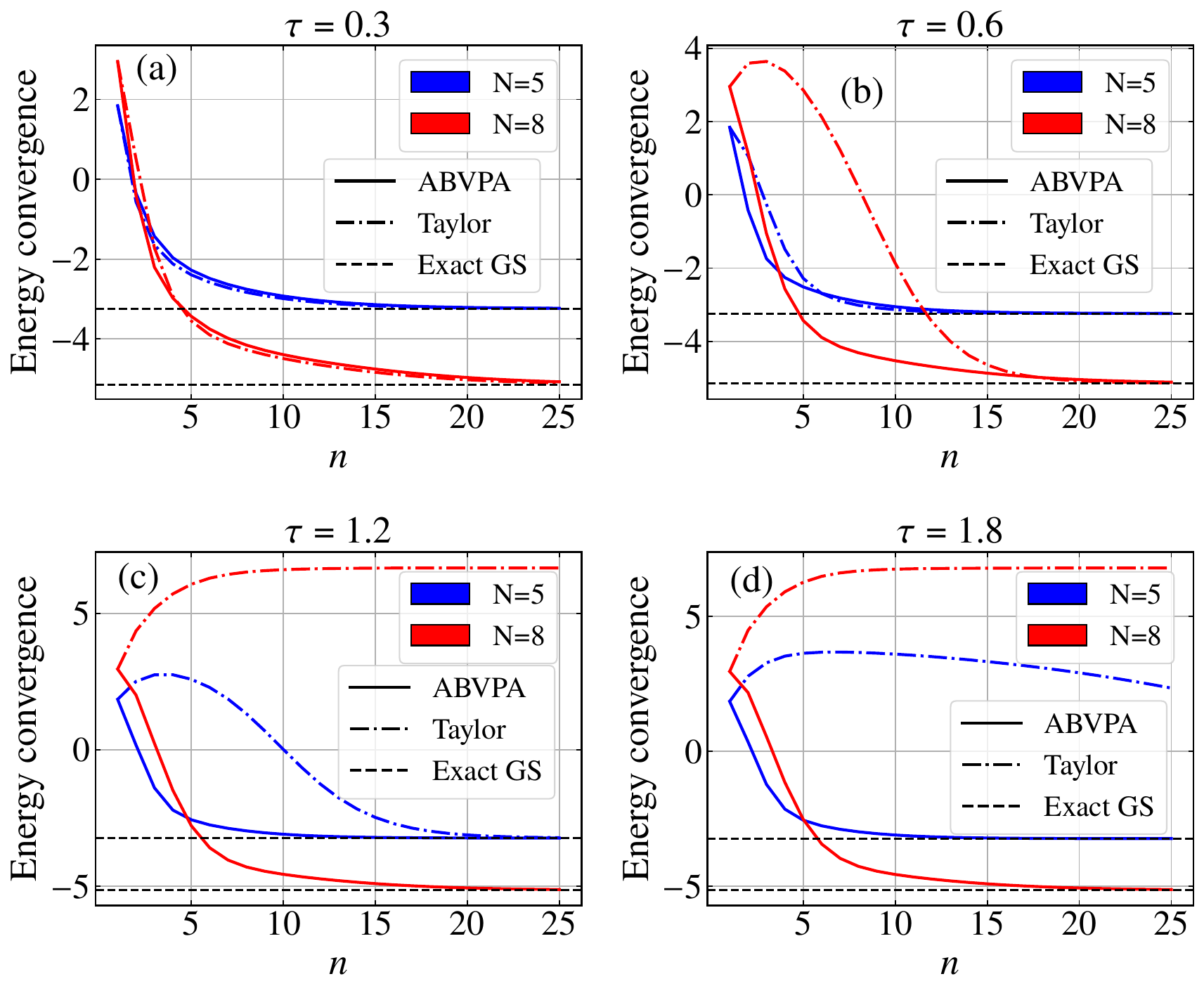}
    \caption{Convergence of the ground-state energy for the QIM (Eq.\eqref{eqn:QIM}) as a function of the number of time steps $n$, comparing the fixed-coefficient Taylor projector with its variational generalization, the imaginary time ABVPA, for (a) $\tau=0.3$, (b) $\tau=0.6$, (c) $\tau=1.2$, and (d) $\tau=1.8$. The Hamiltonian parameters are $J=h_z=h_x=1.0$, for system sizes $N=5$ and $N=8$.}
    \label{fig:Ising_var_Tay_GS}
\end{figure}

Similar to the GS projection using Taylor expansion, one can use our improved expansion of the ITE operator to obtain the GS of an arbitrary qubits Hamiltonian by repeatedly applying $c_0'(\tau)+c_1'(\tau)H$ to a random initial state $\ket{\psi_0}$. After sufficient $n$ repeated applications, one obtains
\begin{equation}
\ket{E_0}\approx\mathcal{N}\left[c_0'(\tau)+c_1'(\tau)H\right]^n\ket{\psi_0},
\end{equation}
where $\mathcal{N}$ is a normalization constant and $n$ is a positive integer. As an illustration, we consider the QIM and plot the convergence of the GS energy as a function of $n$, the number of steps, at different $\tau$ values and with $N=5$ and $N=8$ qubits as shown in Fig. \ref{fig:Ising_var_Tay_GS}. At small $\tau$, the convergence between the two projectors is comparable as shown in Fig. \ref{fig:Ising_var_Tay_GS} (a,b), while at larger $\tau$ we observe that the imaginary-time \mbox{ABVPA} is still valid and stable like the case with $N=8$, $\tau=1.2$ in Fig. \ref{fig:Ising_var_Tay_GS} (c) where the Taylor expansion failed after we increased the number of qubits from $N=5$ to $N=8$. Lastly, the case with $\tau=1.8$ shows that Taylor approximation failed to yield the GS after $25$ steps meanwhile the ABVPA sustained its validity as shown in Fig. \ref{fig:Ising_var_Tay_GS} (d).    

The other Taylor approximation we used was for the real-time evolution $e^{-itH}=I-itH$, which we can replace using action principle \cite{Vogl2025-ku} (See Appendix \ref{app:S_real_TE}). Or alternatively, the parameters in the approximate ABVPA, $e^{-itH}=c_0(t)+c_1(t)H$ can be simply obtained by applying Wick rotation $it\to\tau$ to Eq. \eqref{eqn:cs_imaginaryTE}, giving
\begin{equation}
\label{eqn:cs_realTE}
\begin{aligned}
c_0(t)& = e^{-i \frac{\mu_3}{2\mu_2} t} \left[ \cos(\Omega t) + i \frac{\mu_3}{2\mu_2\Omega} \sin(\Omega t) \right],\\
c_1(t) &= -i\, e^{-i \frac{\mu_3}{2\mu_2} t} \frac{\sin(\Omega t)}{\Omega}.
\end{aligned}
\end{equation}

Lastly, motivated by the recent work in Ref. \cite{Assi2026}, we use alternative approximation to the symmetric second order TS formula that takes the form
\begin{equation}
\label{eqn:TS_var_ansatz}
U_{\rm TS}^{\rm var}(t)=e^{id_0(t)A}e^{id_1(t)B}e^{id_2(t)A}.
\end{equation}
where the variational parameters $d_0(t)$, $d_1(t)$, and $d_2(t)$ are obtained by solving Eq.~\eqref{eqn:ds_for_var_TS} subject to the initial conditions $d_0(0)=d_1(0)=d_2(0)=0$ \cite{Assi2026}.

For short times, it was found that the variational parameters in the above ansatz can be approximated to be \cite{Assi2026}
\begin{equation}
\label{eqn:approx_cs_3exp}
\begin{aligned}
    d_{0}(t)&=d_{2}(t)\approx-\frac{t}{2}+\frac{d_{0,2}^{(3)}(0)}{3!}t^3,\\
    d_1(t)&\approx-t+\frac{d_1^{(3)}(0)}{3!}t^3,
\end{aligned}
\end{equation}
where
\begin{equation}
\label{eqn:VTSC}
\begin{aligned}
    d_{0}^{(3)}(0)&=d_{2}^{(3)}(0)=-\chi\frac{\mathrm{Tr}[B^2]+\frac{1}{2}\mathrm{Tr}[AB]}{2},\\
    d_{1}^{(3)}(0)&=\chi\left[\mathrm{Tr}[AB]+\frac{1}{2}\mathrm{Tr}[A^2]\right],
\end{aligned}
\end{equation}
and
\begin{equation}
    \chi=\frac{\mathrm{Tr}[A^2B^2]-\mathrm{Tr}[(AB)^2]}{\mathrm{Tr}[A^2]\mathrm{Tr}[B^2]-(\mathrm{Tr}[AB])^2}.
\end{equation}
For completeness, we have computed the static traces appearing in the above expression for the QIM in Appendix \ref{app:Traces}. 

\begin{figure}
    \centering
    \includegraphics[width=\linewidth]{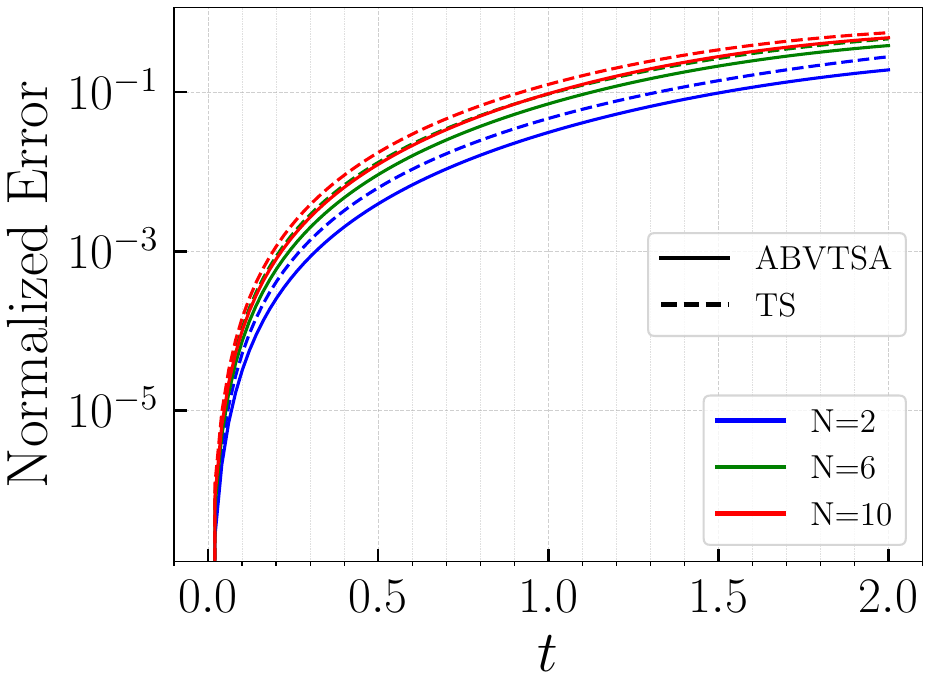}
    \caption{Log-scale plot of the normalized Frobenius-norm error $\frac{1}{2\sqrt{2^N}}\|e^{-itH}-U_{\rm approx}\|_F$ as a function of time $t$ for the QIM (Eq.~\eqref{eqn:QIM}). Here $U_{\rm approx}$ denotes either the fixed-coefficient TS formula (Eq.~\eqref{eqn:TS_formula}) or the ABVTSA (Eq.~\ref{eqn:TS_var_ansatz}). The Hamiltonian parameters are $J=h_z=h_x=1$, for qubit numbers $N=2,6,$ and $10$.}
    \label{fig:Var_vs_TS_vs_N}
\end{figure}

As an illustration, we plot the normalized error between the exact time-evolution operator and the approximate unitary obtained via the the TS decomposition and ABVTSA for the benchmark QIM Eq.~\eqref{eqn:QIM}  at different $N$ values, as shown in Fig.~\ref{fig:Var_vs_TS_vs_N}. Our results clearly demonstrate that the ABVTSA outperforms the TS formula over the entire time interval simulated, and this advantage persists as the system size increases.

\begin{figure}[t]
    \centering
    \includegraphics[width=1\linewidth]{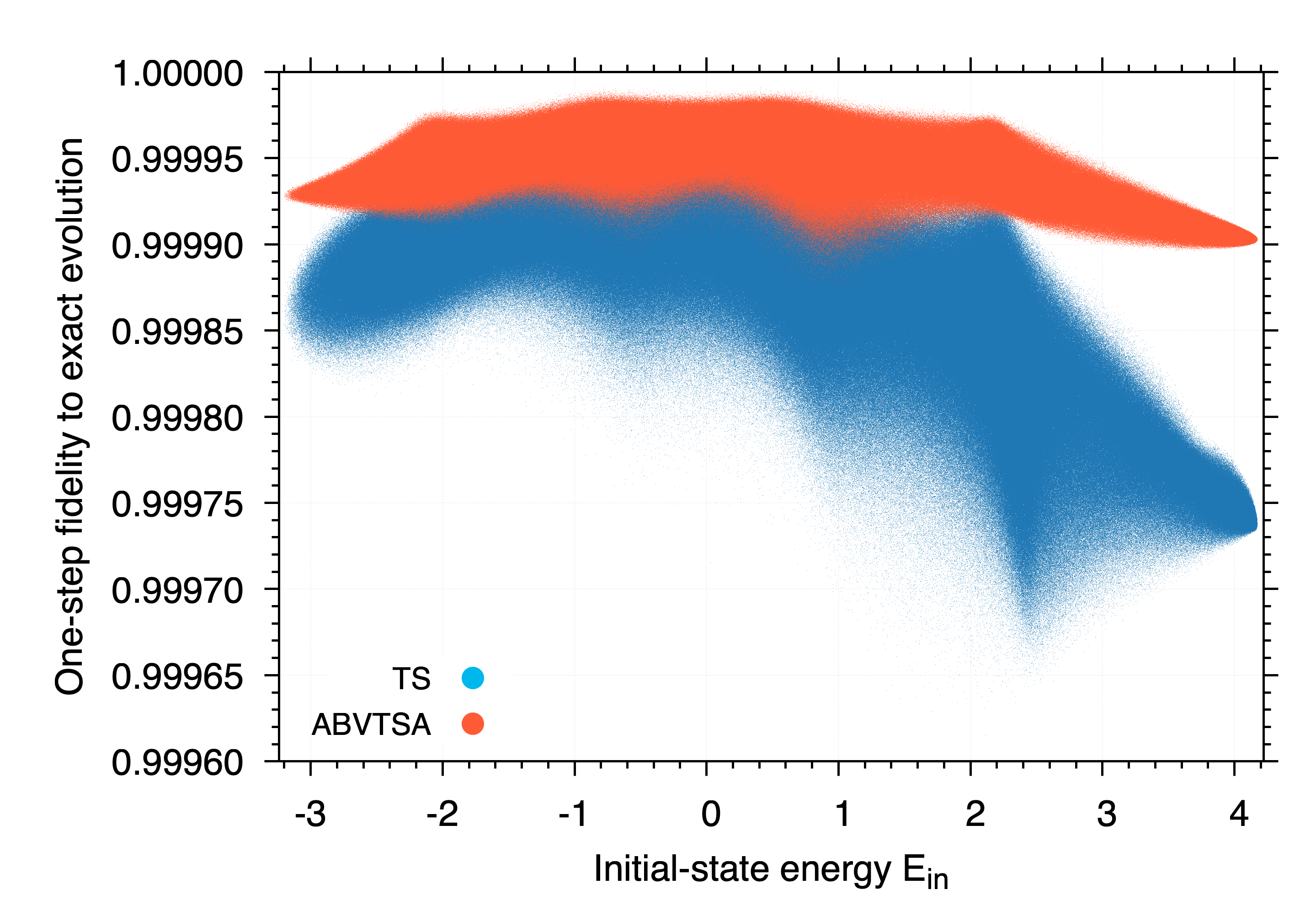}
    \caption{Time-evolution fidelity $F$ as a function of the initial-state energy $E_{\mathrm{in}}=\bra{\psi_{\mathrm{in}}}H\ket{\psi_{\mathrm{in}}}$, comparing the standard TS decomposition with the ABVTSA of Eq.~\eqref{eqn:TS_var_ansatz}. Each point corresponds to one of $10^{7}$ randomly sampled initial states $\ket{\psi_{\mathrm{in}}}$ of the benchmark QIM with $N=5$, evolved for a single time step $\delta t=0.4$. The ABVTSA suppresses the energy-dependent deviations of the standard TS decomposition, yielding a fidelity that remains close to unity.}
    \label{fig:FidelityEnergy}
\end{figure}

In addition, Fig.~\ref{fig:FidelityEnergy} shows the one-step fidelity to exact real-time evolution for the QIM with $N=5$. We sample $10^{7}$ random initial states, denoted by $\ket{\psi_{\text{in}}}$, and characterize each state by its energy expectation value, $E_{\mathrm{in}}=\bra{\psi_{\mathrm{in}}}H\ket{\psi_{\mathrm{in}}}$. For each initial state, we evolve the system for a single time step $\delta t$ using the exact propagator and compare the resulting state with those obtained from the standard TS decomposition and the ABVTSA. The one-step fidelity is defined as $F = \lvert \langle \psi_{\mathrm{exact}}(\delta t) | \psi_{\mathrm{approx}}(\delta t)\rangle \rvert^{2}$. 

For the standard TS decomposition , fidelity depends strongly on $E_{\text{in}}$: states with energies near the extremes of the spectrum show significantly larger deviations. In contrast, the ABVTSA achieves consistently higher fidelity across the entire energy range, and at fixed $E_{\text{in}}$ its fidelity distribution is substantially narrower. These results are shown for $\delta t=0.4$, but we have verified that the same behavior holds for a wide range of physically relevant time steps.

Using the above approximations, one can follow the steps of ancilla-based ITE described in Sec. \ref{sec:ITE-method} to implement these improved expansions as an effective LCU circuit as described in Fig. \ref{fig:ITE_circuit}(e). Thus, the same target-map construction as in Eq.~(\ref{eq:target_map}), the ancilla qubit in Eq.~(\ref{eqn:ancilla-qubit}) is prepared in the state
\begin{equation}
\label{eqn:ancilla-qubit-2}
\ket{\chi}
=
\frac{1}{\sqrt{1+|r|^2}}(\ket{0} + r\ket{1}),
\end{equation}
where $r$ is given by Eq.~(\ref{eqe:ancr}) which unlike Taylor expansion is a function of $\delta t$, the Hamiltonian parameters, and the number of qubits.

As an example we consider the linear ABVPA of QIH~\ref{eqn:QIM}. Using the variational parameters, the ancilla-qubit coefficient $r$ appearing in Eq.~(\ref{eqe:ancr}) for small time step and $\delta t=\delta\tau$ takes the form
\small{
\begin{equation}
r =
-\frac{e^{i\frac{\mu_3}{2\mu_2}\delta t}\sinh(\Omega\delta t)}
{\sinh(\Omega\delta t)\cos(\Omega\delta t)
-i\sin(\Omega\delta t)\cosh(\Omega\delta t)} .
\end{equation}
}
where $\Omega$ and the relevant operator traces $\mu_i$ are the same quantities defined in Eq.~(\ref{eqn:cs_realTE}). The resulting ratio $r$ determines the ancilla-state preparation in Eq.~(\ref{eqn:ancilla-qubit}), which can be implemented using an $R_y(\phi)$ rotation with $\phi=2\arctan(|r|)$ followed by an $R_z(\theta)$ rotation, where $\theta=\arg(r)$. This ABVA implementation is represented in Fig.~\ref{fig:ITE_circuit}(e).

The effect of the ABVA construction on the ITE dynamics is shown in
Fig.~\ref{fig:ITE_circuit}(d), together with the TS--Taylor result introduced above. This panel is intended to illustrate the mechanism of the ITE procedure through a single realization of the quantum evolution rather than to provide a statistical characterization of the energy convergence. Both approaches initially drive the energy toward the GS value; however, the ITE corresponding to the ABVA approaches it more smoothly and remains close to the GS energy at later steps. In contrast, the TS--Taylor approximation eventually deviates from the exact ITE trajectory and overshoots the GS energy.
This behavior reflects the accumulation of approximation errors over successive ITE steps and indicates the improved stability provided by the optimized ABVA parameters.

A quantitative comparison of the convergence behavior is presented later in the results section, where the energy is estimated using a larger number of quantum measurements and the associated statistical uncertainty is explicitly evaluated. In that analysis, the fluctuations and loss of stability of the TS--Taylor evolution become more clearly resolved, allowing a more systematic comparison with the ABVA approach.

\section{Error bounds and resource requirements}
\label{sec:error}
In the ancilla-based ITE procedure, the time step $\delta\tau$ controls both the accuracy and computational cost of the simulation. A smaller time step improves the short-time approximations but requires more ITE steps to reach the same total imaginary time, increasing circuit depth, gate counts, and ancilla measurements. Since each step also requires successful postselection, increasing the number of steps reduces the overall success probability. Conversely, a larger time step reduces the number of ITE steps and circuit resources but increases the approximation errors. The aim of this section is therefore to identify a practical time-step regime that balances accuracy, success probability, and resource requirements.

Two different approximation errors enter the ancilla-based ITE implementation. The first arises from the linear Taylor or ABVPA used to construct the postselected filtering operator, while the second arises from the TS or ABVTSA decomposition of the real-time unitary blocks used within the ITE circuit. Our goal is to derive state-independent error bounds that can be evaluated from properties of the Hamiltonian using only simple calculations, without requiring explicit knowledge or preparation of a particular quantum state. These bounds therefore provide a classical pre-computation procedure for selecting an appropriate time step directly from Hamiltonian information, before executing the quantum circuit. We first analyze the two error contributions and their characteristic time scales, and then examine the qubit and gate requirements of the implementation.

We now proceed into deriving error bounds for our approximations that will be useful in identifying the proper time steps. First, we recall that the exact time-evolution operator $U_{\rm ex}(t)=e^{-itH}$ satisfies $i\dot{U}_{\rm ex}(t)=HU_{\rm ex}(t)$, we define the global error of an approximate ansatz $U_a(t)$ as
\begin{equation}
    \epsilon(t)=\frac{\|\Delta(t)\|_F}{2\sqrt{D}},
\end{equation}
where $\|\cdot\|_F$ denotes the Frobenius norm, $\Delta(t)=U_{\rm ex}(t)-U_a(t)$, and $D$ is the Hilbert-space dimension. For a unitary ansatz, $\epsilon(t)$ takes values between 0 and 1. Noting that
\begin{equation}
    \dot{\Delta}(t)+iH\Delta(t)=iR(t),\qquad \Delta(0)=0,
\end{equation}

where $R(t)=i\dot{U}_a(t)-HU_a(t)$. Multiplying both sides by $e^{itH}$, the above equation can be rewritten as
\begin{equation}
    \frac{d}{dt}\left[e^{itH}\Delta(t)\right]=ie^{itH}R(t),
\end{equation}
integrating both sides, we obtain
\begin{equation}
    \Delta(t)=i\int_0^t ds\,e^{i(s-t)H}R(s).
\end{equation}
Using the triangle inequality, we obtain the following upper bound on the global error
\begin{equation}
\label{eqn:Epsilons}
    \epsilon(t)=\frac{1}{2\sqrt{D}}\left\|i\int_0^t ds\,e^{i(s-t)H}R(s)\right\|_F\leq \epsilon^*(t),
\end{equation}
where $\epsilon^*(t)$ is the error bound given below
\begin{equation}
\label{eqn:eps_star}
    \epsilon^*(t)=\frac{1}{2\sqrt{D}} \int_0^t ds\,\left\|R(s)\right\|_F .
\end{equation}

For Taylor expansion $U_a(t)=1-itH$, we find $\left\|R(s)\right\|_F=t\left\|H^2\right\|_F$, giving $\epsilon^*(t)_{\text{Taylor}}=\frac{t^2}{4\sqrt{D}}\left\|H^2\right\|_F$. The rule of thumb is to choose the time step $\delta t$ so that we have a small error bound $\epsilon(\delta t)$, or alternatively, $\delta t<t^*_{\text{Taylor}}=2D^{1/4}/\sqrt{\left|H^2\right\|_F}$. Similarly, we derive an expression for the error bound for the ABVPA, $e^{-itH}\approx c_0(t)+c_1(t)H$. First, we obtain the following formula for $||R||_F^2$

\begin{equation}
\label{eqn:R2}
    ||R||_F^2=\left(\mu_4-\frac{\mu_2^2}{\mu_0}-\frac{\mu_3^2}{\mu_2}\right)\frac{\sin^2(\Omega t)}{\Omega^2}.
\end{equation}
Using Eq. \eqref{eqn:eps_star}, we find
\begin{equation}
\label{eqn:EpsilonsStar}
    \epsilon^*(t)_{\text{ABVPA}}=\frac{1}{2\sqrt{D}} \sqrt{\mu_4-\frac{\mu_2^2}{\mu_0}-\frac{\mu_3^2}{\mu_2}}\left[\frac{1-\cos(\Omega t)}{\Omega^2}\right].
\end{equation}
These equations are helpful in determining the acceptable time steps so that we maintain errors below a given bound. 

Using Taylor expansion $1-\cos(\Omega t)\approx\frac{\Omega^2 t^2}{2}+\mathcal{O}(t^4)$, thus, we can approximate the error bound as
\begin{equation}
     \epsilon^*(t)_{\text{ABVPA}}\approx \left(\frac{t}{t^*_{\text{ABVPA}}}\right)^2,
\end{equation}
where
\begin{equation}
\label{eqn:t_start}
    t^*_{\text{ABVPA}}=2\sqrt{\frac{D}{\mu_4-\frac{\mu_2^2}{\mu_0}-\frac{\mu_3^2}{\mu_2}}},
\end{equation}
defines a characteristic time scale for the validity of the linear approximation in ABVA. For the benchmark QIM~\ref{eqn:QIM}, with $N=5$, we obtain $t^*_{\text{ABVPA}}=0.568$. We then choose the simulation time step $\delta t=0.4$, which lies below this characteristic scale. Evaluating the global error at this time step gives $\epsilon_{\text{ABVPA}}=0.130$, while the corresponding error bound is $\epsilon^*_{\text{ABVPA}}=0.136$. These values indicate that the chosen time step remains within the regime where the linear approximation is accurate. For comparison, applying the same analysis to the first-order Taylor approximation gives $t^*_{\text{Taylor}}=0.937$, $\epsilon_{\text{Taylor}}=0.173$, and $\epsilon^*_{\text{Taylor}}=0.182$ for the same Hamiltonian parameters

We also evaluate the errors associated with the  TS decomposition and its action-based variational extension, ABVTSA. The detailed derivations provided in Appendix~\ref{app:TS_errors}. Since both constructions reproduce the exact propagator through second order in $t$, their leading errors scale as $t^3$ rather than $t^2$:
\small{
\begin{equation}
    \epsilon_{\rm TS}^*(t)\simeq
    \left(\frac{t}{t_{\rm TS}^*}\right)^3,
    \quad
    \epsilon_{\rm ABVTS}^*(t)\simeq
    \left(\frac{t}{t_{\rm ABVTSA}^*}\right)^3.
\end{equation}
}
where $t_{\rm TS}^*$ and $t_{\rm ABVTSA}^*$ are the characteristic time scales that determine the short-time validity of the corresponding approximations.

For the benchmark QIM in Eq.~\eqref{eqn:QIM}, with $N=5$,, the characteristic time scales obtained from the analysis in Appendix~\ref{app:TS_errors} are $t_{\rm TS}^*=2.252$ and $t_{\rm ABVTS}^*=2.625$. At the simulation time step $\delta t=0.4$, direct numerical evaluation of the propagator errors and residual-based bounds gives $\epsilon_{\rm TS}=0.0054$ and $\epsilon_{\rm TS}^*=0.00553$ for the standard TS decomposition, and $\epsilon_{\mathrm{ABVTSA}}=0.00349$ and $\epsilon_{\mathrm{ABVTSA}}^*=0.00353$ for ABVTSA. These values show that $\delta t=0.4$ lies well within the short-time regime of both decompositions, with errors below approximately $0.6\%$. The variational construction further reduces both the global error and its upper bound relative to the standard TS decomposition.

Thus, at the chosen time step, the real-time decomposition error remains substantially smaller than the linear filtering error, making the ancilla-based filtering approximation the primary accuracy constraint on the step size. The effect of the time step on the postselection probability can be understood from Eq.~\eqref{eqn:Err2}. For any state, $\langle H^2\rangle_\psi\geq 0$, so the quadratic contribution $\langle H^2\rangle_\psi\delta\tau^2/6$ is nonnegative and increases with positive $\delta\tau$. Moreover, for any nonzero traceless Hermitian Hamiltonian, including our benchmark QIM with $\mathrm{Tr}(H)=0$, the ground-state energy satisfies $E_{\rm GS}<0$. Hence, as the evolved state approaches the GS, $\langle H\rangle_\psi\simeq E_{\rm GS}<0$, making the linear contribution $-\langle H\rangle_\psi\delta\tau/3$ positive as well. Therefore, for a state sufficiently close to the GS, both corrections in Eq.~\eqref{eqn:Err2} increase the per-step postselection probability as $\delta\tau$ increases. Within the regime of validity of the small-$\delta\tau$ expansion, $p_{\rm step}$ is therefore locally increasing with the time step. A larger admissible time step also reduces the number of ITE steps required to reach a fixed total imaginary time, further improving the overall postselection success probability.

We next consider the hardware resources required for the implementation in terms of qubit and gate counts. The qubit requirements depend on the system size and the number of ancilla qubits used to implement the measurement-induced imaginary-time evolution steps. We consider an $N$-qubit system and $n$ imaginary-time steps. In the simplest circuit implementation, each ITE step requires a fresh ancilla qubit that is measured at the end of the step and then discarded. Consequently, the total number of qubits is $N_{\mathrm{tot}}=N+n$, and the number of ancilla measurements is $n$. If the hardware supports mid-circuit measurement and qubit reuse, the same ancilla qubit can be reused for all ITE steps~\cite{decross2023qubit,hua2023caqr}. In that case, the total qubit count reduces to $N_{\mathrm{tot}}=N+1$, while the number of ancilla measurements remains $n$.

The gate-count comparison is reported for the Hamiltonian in Eq.~\eqref{eqn:QIM} and its decomposition in Eq.~\eqref{eqn:HamiltonianParts}. For an open chain, the Hamiltonian contains $N-1$ nearest-neighbor $S_j^zS_{j+1}^z$ contributions, $N$ single-qubit $S_j^x$ contributions, and $N$ single-qubit $S_j^z$ contributions. The corresponding logical and backend-transpiled gate counts for $N=5$ are summarized in Table~\ref{tab:gate_counts}.

\begin{table}[t]
\centering
\small
\begin{tabular}{llll}
\hline
Circuit  &  Logical  &   Backend  \\
\,  &   gates &    gates \\
\hline
Hamiltonian terms & $4N-2 = 18$ & 18 (Analytic)  \\
\hline
TE step: $U(\delta t)$ & $5N-2 = 23$ & 18 \\
ITE step: $\Lambda(U(\delta t))$ & $5N-2 = 23$ & 123 \\
\hline
ABVA  TE step: $U(\delta t)$ & $5N-2 = 23$ & 18 \\
ABVA ITE step: $\Lambda(U(\delta t))$ & $5N-2 = 23$ & 123 \\
\hline
\end{tabular}
\caption{
Gate-count comparison for the QIM with $N=5$. Logical-gate counts are obtained analytically before transpilation. Aer-transpiled counts were obtained using Qiskit 2.1.1 and Qiskit Aer 0.17.1 with the \texttt{aer\_simulator\_matrix\_product\_state} backend. For the Hamiltonian row, the count corresponds to the analytic term-by-term Pauli-gate implementation.  The ABVA modifies the circuit parameters without changing its structure and therefore requires the same number of gates as the corresponding nonvariational TE and ITE circuits.
}
\label{tab:gate_counts}
\end{table}

For the term-by-term Hamiltonian implementation, each $S_j^zS_{j+1}^z$ contribution is counted as two Pauli $Z$ gates, each $S_j^x$ contribution as one Pauli $X$ gate, and each $S_j^z$ contribution as one Pauli $Z$ gate. This gives a logical count of $2(N-1)+N+N = 4N-2$, which is 18 gates for $N=5$.

For the time evolution circuit, one TS step implements the unitary $U(\delta t)$ according to Eq.~\eqref{eqn:TS_formula}, using the decomposition in Eq.~\eqref{eqn:HamiltonianParts}. Since $A$ contains the $S_j^zS_{j+1}^z$ and $S_j^z$ terms, while $B$ contains the $S_j^x$ terms, one TE step requires $2(N-1)$ $R_{ZZ}$ rotations, $2N$ $R_Z$ rotations, and $N$ $R_X$ rotations, giving $2(N-1)+2N+N = 5N-2$ logical gates, or 23 logical gates for $N=5$. The direct ITE step uses the controlled unitary $\Lambda(U(\delta t))$, which has the same logical structure as $U(\delta t)$, except that the corresponding rotations are controlled by the ancilla. Therefore, one direct $\Lambda(U)$ in ITE step also contains $5N-2$ logical controlled rotations before transpilation. Although time evolution and ITE have the same logical count, their backend-transpiled gate counts differ substantially because the controlled rotations in $\Lambda(U)$ are decomposed into backend-supported gates. For $N=5$, the TE step gives 18 gates after transpilation, while the direct controlled-$U_a$ ITE step gives 123 gates.

The ABVA implementation retains the same TS circuit structure and differs only in the values of the optimized evolution parameters. It therefore introduces no additional gates relative to the corresponding standard TS implementation. Consequently, the ABVA time evolution and ITE steps have the same logical and backend-transpiled gate counts as their standard counterparts, namely 23 logical gates for both cases and 18 and 123 backend gates for time evolution and ITE, respectively, as summarized in Table~\ref{tab:gate_counts}.

\section{Results and Discussions}
\label{sec:result}
\begin{figure}[t]
    \centering
    \hspace{10cm}
    \includegraphics[width=1\linewidth]{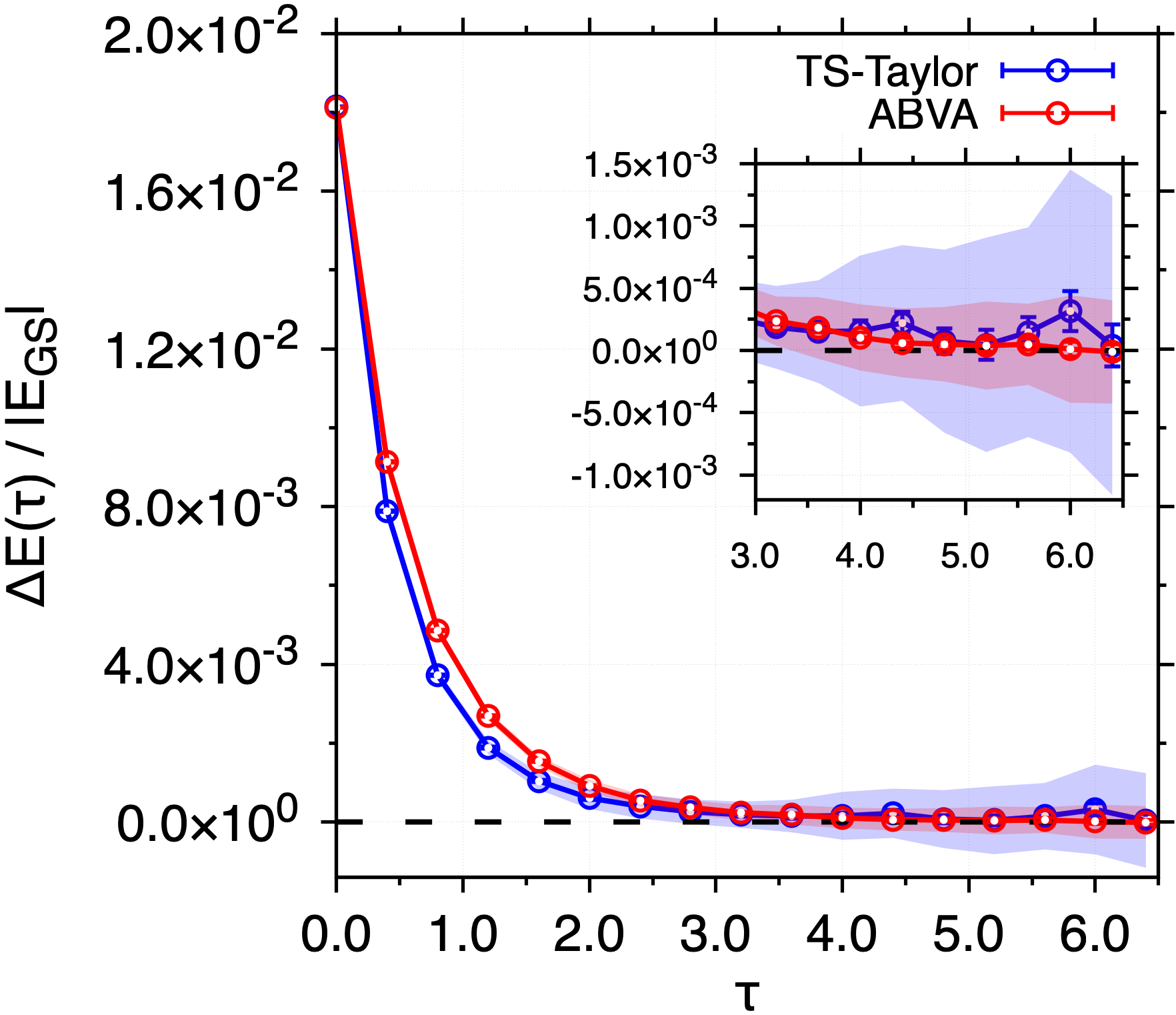}
    \caption{Convergence of the relative energy deviation $\Delta E(\tau)/|E_{\mathrm{GS}}|$ during ITE, with $\Delta E(\tau)=E(\tau)-E_{\mathrm{GS}}$. Here, $E_{\mathrm{GS}}$ denotes the exact GS energy of the five-qubit benchmark QIM. The TS--Taylor ITE method (blue) and the ABVA implementation for ITE (red) both approach the exact GS energy as $\tau$ increases, using a time-step size of $\delta\tau=0.4$. Shaded bands show the standard deviation over 50 independent runs, each using $10^7$ measurement shots. The inset magnifies the late-time interval and shows the residual statistical fluctuations near convergence.}
    \label{fig:Energy}
\end{figure}

\subsection{Energy convergence and statistical stability}
Figure~\ref{fig:Energy} benchmarks the TS--Taylor and ABVA, ITE implementations for the five-qubit QIM. To enable a controlled comparison of convergence to the exact GS, the evolution is initialized from a mean-field (MF) product-state approximation. In the MF treatment, the interaction is decoupled according to
\begin{equation}
S_i S_{i+1} \approx S_i \langle S_{i+1} \rangle 
+ \langle S_i \rangle S_{i+1} 
- \langle S_i \rangle \langle S_{i+1} \rangle.
\end{equation}

This approximation reduces the many-body problem to an effective single-qubit description. The GS of the resulting mean-field Hamiltonian is then used to construct a physically motivated product-state initial condition for the subsequent ITE.

As shown in Fig.~\ref{fig:Energy}, both TS--Taylor and ABVA implementations rapidly reduce the relative energy deviation $\Delta E(\tau)/|E_{\mathrm{GS}}|$ at small $\tau$, consistent with the exponential suppression of excited-state components under imaginary-time propagation. For increasing $\tau$, the estimated energy approaches $E_{\mathrm{GS}}$ and saturates at a near-zero relative error, indicating successful GS projection. The inset ($3\le\tau\le6$) resolves the late-time regime, where residual fluctuations reflect finite sampling and accumulated numerical errors. 

To quantify statistical stability, Fig.~\ref{fig:Energy} also reports the standard deviation (SD) of the measured energy versus $\tau$. While both approaches show comparable uncertainty at small $\tau$, the TS--Taylor implementation exhibits a more pronounced increase in SD at larger $\tau$, indicating increased sensitivity to approximation and sampling errors. In contrast, the ABVA maintains a lower SD in the long-time regime, demonstrating improved numerical robustness.

\begin{figure}[t]
    \centering
    \includegraphics[width=1\linewidth]{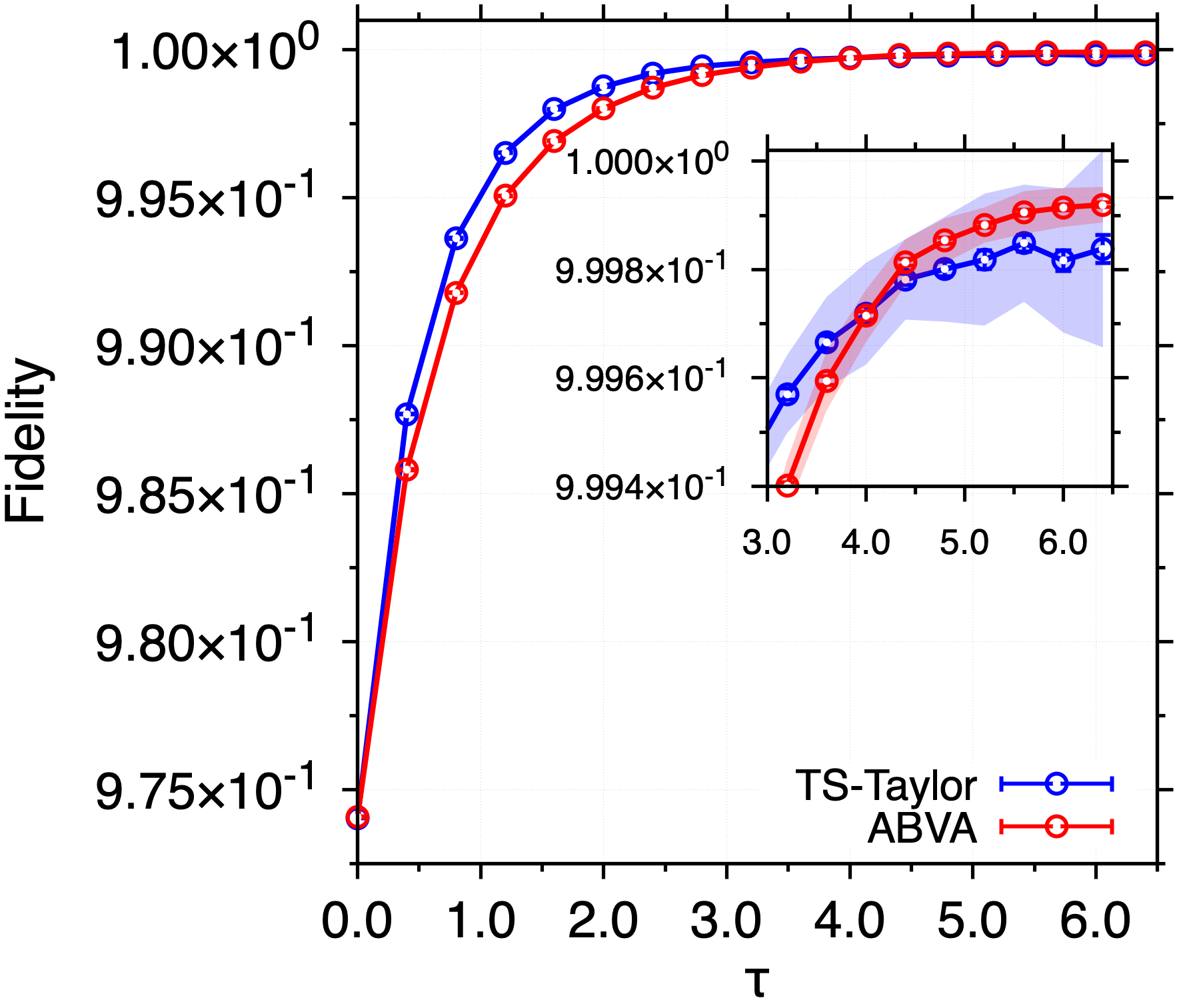}
    \caption{GS fidelity as a function of imaginary time $\tau$ for the TS--Taylor ITE (blue) and ABVA ITE (red), using the same simulation setup as in Fig.~\ref{fig:Energy}. Both methods converge toward unit fidelity, with ABVA showing improved late-time fidelity and reduced fluctuations. Shaded bands represent the standard deviation over independent runs, and the inset magnifies the late-time behavior near convergence.
}
    \label{fig:Fidelity}
\end{figure}

\subsection{Fidelity convergence}
Figure~\ref{fig:Fidelity} reports the fidelity with respect to the exact GS for the same simulation setup as in Fig.~\ref{fig:Energy}. Both implementations yield a rapid increase in fidelity at small $\tau$, followed by saturation close to unity for $\tau\gtrsim 3$, confirming successful GS preparation. The inset highlights small systematic differences near convergence and also shows the corresponding standard deviation of the fidelity.

While both methods exhibit similar fluctuations at early times, the TS--Taylor implementation develops slightly larger variability at longer imaginary times $\tau$, whereas the variational approach maintains a consistently smaller SD. Notably, a reduction in fidelity is observed near the final time steps for the standard method.

This behavior can be understood from Fig.~\ref{fig:FidelityEnergy}, which shows that the distribution of fidelity values is more concentrated for the ABVA, particularly in the low-energy region where the state approaches the GS during ITE. As the system evolves toward lower energies at large $\tau$, the narrower and higher-fidelity distribution of the ABVTSA leads to reduced fluctuations and improved stability compared to the standard TS scheme. Together with the energy analysis, these results indicate that the ABVA provides improved stability in the long-time regime.

\begin{figure}[t]
    \centering
    \includegraphics[width=0.9\linewidth]{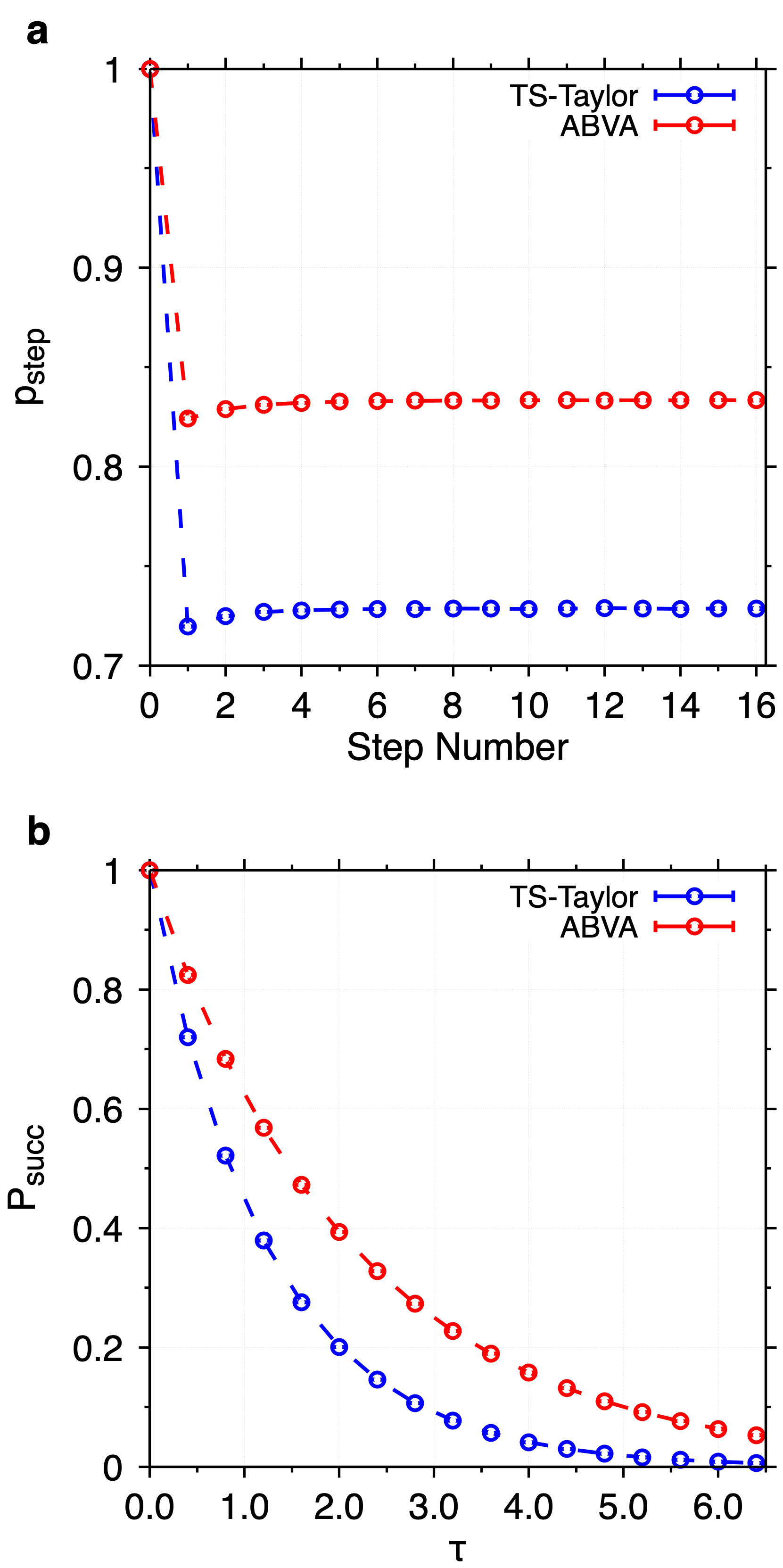}
    \caption{
(a) Per-step success probability, $p_{\mathrm{step}}$, as a function of the algorithmic step number for the TS--Taylor ITE (blue) and ABVA ITE (red). (b) Total success probability, $P_{\mathrm{succ}}$, as a function of imaginary time $\tau$. Both panels use the same simulation setup and imaginary-time step size, $\delta\tau=0.4$, as Fig.~\ref{fig:Energy}. At $\tau=0$, no postselection has yet been performed, so all shots are retained and $p_{\mathrm{step}}=P_{\mathrm{succ}}=1$. After the first ITE step, both methods approach an approximately constant per-step success probability, while the cumulative success probability decreases with increasing $\tau$ due to the probabilistic nature of the evolution. Error bars represent the standard error of the mean.
}
    \label{fig:Success}
\end{figure}

\subsection{Success probability}
\label{sec:success_results}

Figure~\ref{fig:Success} characterizes the probabilistic performance of the ancilla-assisted implementations. Figure~\ref{fig:Success}(a) shows the per-step success probability versus algorithmic step number. After the initial transient, both methods reach an approximately steady per-step success rate, with the ABVA exhibiting a consistently higher success probability, indicating a more reliable update at each imaginary-time step.

The cumulative success probability is shown in Fig.~\ref{fig:Success}(b) as a function of $\tau$. As expected for a post-selected protocol, the total success probability decreases with increasing $\tau$ due to the multiplicative accumulation of per-step success probabilities. The TS--Taylor implementation exhibits a faster decay, consistent with its lower per-step success rate. In contrast, the ABVA retains a substantially higher cumulative success probability across the full evolution. Error bars in Fig.~\ref{fig:Success} represent the standard error of the mean.

By replacing the fixed Taylor coefficients and standard TS coefficients with optimized action-based variational parameters at the operator level, the proposed method improves the accuracy of each imaginary-time step and increases the per-step success probability relative to the corresponding fixed-coefficient implementation, all without increasing circuit depth. In the benchmark cases studied here, this preserves a compact probabilistic ITE circuit while improving the effective imaginary-time filter, leading to higher fidelity and enhanced success probability at comparable depth.

Overall, the combined energy, fidelity, and success-probability benchmarks demonstrate that the variational construction enhances the accuracy of each filtering step while also increasing the post-selection success probability relative to the fixed-coefficient TS-Taylor implementation. This balance between approximation accuracy and probabilistic success is central to making ITE-based GS preparation more practical on near-term quantum devices.

\section{Conclusions}
\label{sec:conclusions}
In this work, we developed a variational refinement of ancilla-assisted probabilistic ITE for GS preparation. Starting from a single ancilla ITE circuit based on a first-order Taylor approximation and a second-order Trotter-Suzuki decomposition, we replaced the fixed coefficients with variationally optimized parameters at the operator level. This construction preserves the basic circuit architecture while improving the effective imaginary-time filter, and it requires only a single ancillary qubit per full imaginary-time step, substantially reducing the ancillary overhead compared with term-wise probabilistic implementations.

The proposed method occupies a distinct point in the resource landscape. Unlike VQE, it does not rely on a restricted state ansatz or suffer from increasingly challenging classical optimization over circuit parameters; instead, it optimizes the propagator approximation itself~\cite{Peruzzo2014-yg,TangPRXQuantum2021,AnastasiouPhysRevResearch2024}. Compared with other gate-based probabilistic ITE schemes that assign ancillary qubits to individual Hamiltonian terms, our single-ancilla construction simplifies post-selection, reduces the required Hilbert-space extension, and mitigates measurement overhead~\cite{liu2021probabilistic,kosugi2022imaginary,wen2023iteration,yi2025probabilistic}.

Looking toward scalability, we note that the optimized coefficients in the linear-polynomial ansatz are determined by trace moments of the Hamiltonian, $\mathrm{Tr}[H^n]$. While the direct evaluation of these quantities may become computationally demanding for generic many-body systems, efficient classical approximations are available for broad classes of models. In particular, tensor-network representations such as matrix product operators (MPO) or tensor trains enable powers of the Hamiltonian to be constructed through successive MPO multiplications, allowing trace moments $\mathrm{Tr}(H^n)$ to be computed with controllable accuracy \cite{PhysRevB.95.035129,doi:10.1137/090752286}. This suggests that the computational cost is largely dictated by the efficiency with which the Hamiltonian can be represented, rather than by the variational optimization itself.

Although the present demonstration focused on the linear polynomial approximation, where $e^{-\tau H}\approx c_0'(\tau)+c_1'(\tau)H$ and the unitary is expressed as $e^{-i\delta t H}\approx c_0(\delta t)+c_1(\delta t)H$ to maintain a single-ancilla LCU (linear combination of unitaries), the variational framework is straightforward to generalize to higher-order expansions. In particular, one may adopt the general polynomial forms $e^{-\tau H}\approx\sum_{j=0}^{k} c_j(\tau) H^j$ and $e^{-i\delta t H}\approx \sum_{j=0}^{m} d_j(\delta t) H^j$, allowing the imaginary-time propagator to be represented as an LCU with systematically improved accuracy. Such higher-order approximations can reduce the number of time steps required for a fixed total imaginary time or enhance fidelity for systems with small spectral gaps. This improvement, however, comes at the cost of requiring additional ancillary qubits to coherently encode the larger number of unitary terms, illustrating a direct resource trade-off between approximation accuracy and circuit complexity within our variational framework.

Despite these advantages, the method remains subject to the inherent trade-offs of probabilistic ITE. The present linear (or low-order) polynomial approximation is effective for the benchmark systems studied here, but more complex Hamiltonians or systems with small spectral gaps may require longer evolution, smaller time steps, or higher-order approximations, each of which can increase circuit complexity or reduce the total post-selection probability. The variational construction helps improve this balance by increasing per-step accuracy and stability without increasing circuit depth relative to the fixed-coefficient TS-Taylor circuit. Additionally, while the variational formulation offers a guideline for estimating stable $\delta t$ ranges, the final choice of imaginary-time step must still be refined by balancing approximation accuracy, fidelity, and post-selection success for the specific Hamiltonian and initial state.

Overall, these results indicate that variationally optimized probabilistic ITE offers a promising and resource-efficient route toward higher-fidelity GS preparation on near-term quantum devices, effectively bridging rigorous energy-filtering methods with practical hardware constraints.

\section{Acknowledgments}
B. seifi, I. Assi and J. P. F. LeBlanc acknowledge the support of the Natural Sciences and Engineering Research Council of Canada (NSERC) RGPIN-2022-03882 and (NRC) AQC-200-1. We thank Meenu Kumari for useful
discussions.

\normalsize
\appendix
\begin{widetext}

\section{Analytic traces for the QIM}
\label{app:Traces}
In this section, we will obtain various static traces related to the QIM that appeared in the real and imaginary time variational principles. We begin by recalling the following convention
\begin{equation}
    \sigma_j^\alpha:=I \otimes \cdots \otimes \underbrace{\sigma^\alpha}_{j\text{th site}} \otimes \cdots \otimes I,
\end{equation}
where $\alpha=x,y$, and $z$. Also, we introduce the following identity
\begin{equation}
    \mathrm{Tr}[V\otimes W]=
    \mathrm{Tr}[V]\mathrm{Tr}[W].
\end{equation}
For the QIM, we can express the Hamiltonian as
\begin{equation}
    H=\sum_{j=1}^{3N-1}w_j P_j,
\end{equation}
where 
\begin{equation}
P_j = 
\begin{cases}
\sigma_j^z \sigma_{j+1}^z, & 1 \le j \le N-1, \\
\sigma_j^z,                  & N \le j \le 2N-1, \\
\sigma_j^x,                  & 2N \le j \le 3N-1,
\end{cases}
\end{equation}
and
\begin{equation}
w_j= 
\begin{cases}
\frac{J}{4}, & 1 \le j \le N-1, \\
\frac{h_z}{2},                  & N \le j \le 2N-1, \\
\frac{h_x}{2},                  & 2N \le j \le 3N-1.
\end{cases}
\end{equation}
Clearly, one has $\mathrm{Tr}\left[P_jP_k\right]=2^N\delta_{jk}$. Using Eq., one automatically finds $\mathrm{Tr}[H]=0$. Similarly, using this orthogonality relation we find
\begin{equation}
    \mathrm{Tr}[H^2]=\sum_{j,k}w_jw_k\mathrm{Tr}\left[P_jP_k\right]=2^N\left[\frac{J_z^2 (N-1)}{16} + \frac{N(h_x^2 + h_z^2)}{4}\right],
\end{equation}

\begin{equation}
    \mathrm{Tr}[H^3]=\sum_{j,k,\ell}w_jw_kw_\ell\mathrm{Tr}\left[P_jP_kP_\ell\right]=2^N\frac{3}{8} J_z h_z^2 (N-1),
\end{equation}

\begin{equation}
\label{eq:TrH4}
\begin{aligned}
\mathrm{Tr}\!\left[H^4\right]
&= \sum_{j,k,\ell,m}w_jw_kw_\ell w_m\mathrm{Tr}\left[P_jP_kP_\ell P_m\right]=2^N \Bigg[
\frac{J^4}{256}\left(3N^2-8N+5\right)
+ \frac{h_z^4}{16}N(3N-2)
+ \frac{h_x^4}{16}N(3N-2)
\\
&+ \frac{3J^2h_z^2}{32}\left(N^2+3N-8\right)
+ \frac{J^2h_x^2}{32}(N-1)(3N-4)+ \frac{h_z^2h_x^2}{16}\left(6N^2-4N\right)
\Bigg].
\end{aligned}
\end{equation}

Thus, we have provided the analytic formulas for the traces appearing in the variational imaginary time-evolution ansatz for our QIM. Now, we would like to provide the traces that also appear in the real time-evolution ansatz. We beging by defining $A=\frac{J}{4}\sum_{j=1}\sigma_j^z\sigma_{j+1}^z+\frac{h_z}{2}\sum_{j=1}^N\sigma_j^z=\sum_{j=1}^{2N-1}w_jP_j$ and $B=\frac{h_x}{2}\sum_{j=1}^N\sigma_j^x=\sum_{j=2N}^{3N-1}w_jP_j$, we find the following traces

\begin{equation}
\label{eqn:var_p1}
    \mathrm{Tr}[A^2]=\frac{(N-1)2^NJ^2}{16}+\frac{N2^Nh_z^2}{4},
\end{equation}
\begin{equation}
\label{eqn:var_p2}
    \mathrm{Tr}[B^2]=\frac{N 2^Nh_x^2}{4},
\end{equation}
\begin{equation}
\label{eqn:var_p3}
    \mathrm{Tr}[A^2B^2]=\frac{N2^Nh_x^2}{4}\left[(N-1)\frac{J^2}{16}+\frac{Nh_z^2}{4}\right],
\end{equation}
and
\begin{equation}
\label{eqn:var_p4}
    \mathrm{Tr}[(AB)^2]=\frac{2^Nh_x^2}{4}\left[\frac{J^2}{16}(N - 1)(N - 4)+\frac{h_z^2}{4}N(N-2)\right].
\end{equation}

\section{Action-based variational principle for real time-evolution}
\label{app:S_real_TE}
In section \ref{sec:var_ITE}, we have introduced an action principle for the imaginary-time evolution in which we presented alternative polynomial approximations to the ITE operator for arbitrary Hamiltonians. 

In this section, we review alternative action principle that applies to real time-evolution \cite{Vogl2025-ku,Assi2026}
\begin{equation}
\label{eqn:action_S}
    S=\int dt \, \mathrm{Tr} \left[ U^{\dagger} ( i \partial_{t} U - H U ) \right].
\end{equation}
By varying $S$ with respect to the entries of $U(t)$ or $U^\dagger(t)$ will yield the Euler-Lagrange equations that automatically result in the Schrodinger equation $i\partial_tU(t)=HU(t)$. In practice, we use an ansatz for the time-evolution denoted by $U_a(\{\theta_j(t)\})$, where $\{\theta_j(t)\}$ are time-dependent variational parameters, then setting $\delta S=0$ for this ansatz result in the following equations of motion
\begin{equation}
\label{eq:real_t_EOM}
    \sum_{k} g_{jk} \dot{\theta}_k + i F_j = 0,
\end{equation}
where $g_{jk} = \mathrm{Tr} \left[ \left( \frac{\partial U_{a}}{\partial \theta_j} \right)^\dagger \frac{\partial U_{a}}{\partial \theta_k} \right]$ 
is the quantum geometric tensor, and 
$F_j = \mathrm{Tr} \left[ \left( \frac{\partial U_{a}}{\partial \theta_j} \right)^\dagger H U_a \right]$ 
is the generalized force \cite{Vogl2025-ku,Assi2026}. These expressions are identical to those given in Eq. \eqref{eqn:diff_eq_var_parms}. Interestingly, one can easily map the Eq. \eqref{eq:real_t_EOM} to Eq. \eqref{eqn:diff_eq_var_parms} via wick rotation $it\to \tau$. 

In this work, we are interested in two real-time evolution Ansatz. Our first ansatz is the action-based variational polynomial ansatz (ABVPA), defined as a polynomial expansion in $H$ \cite{Vogl2025-ku},
\begin{equation}
    U_a^{P}(\{c_j(t)\})=\sum_{j=0}^{M} c_j(t) H^j,
\end{equation}
where $\{c_j(t)\}$ are found by solving Eq. \eqref{eq:real_t_EOM}, or more explicitly
\begin{equation}
    \sum_{k} \mu^{j+k} \dot{c}_k + i \sum_k \mu^{j+k+1}c_k = 0,
\end{equation}
where $\mu_n=\mathrm{Tr}\left[H^n\right]$, and the initial conditions are $c_0(0)=1$ and $c_{j>0}(0)=0$. 

As an illustration, we compare the average error for a sample of random $10\times 10$ Hermitian matrices as shown in Fig. \ref{fig:error_taylor_vs_var_realTE}. The plot shows the normalized error $\frac{1}{2\sqrt{10}}||e^{-itH}-U_{\rm app}(t)||_F$ where for Taylor $U_{\rm app}(t)=1-itH$ while for the linear order ABVPA  we have $U_{\rm app}(t)=c_0(t)+c_1(t)H$. Clearly, the ABVPA showing lower error across all time scales. From the quantum computing prospective, the implementation of $1-itH$ or $c_0(t)+c_1(t)H$ require the same number of ancillary qubits and the advantage of using the ABVPA is in minimizing the error due to the linear approximation compared to the direct Taylor expansion.
\begin{figure}
    \centering
    \includegraphics[width=0.5 \linewidth]{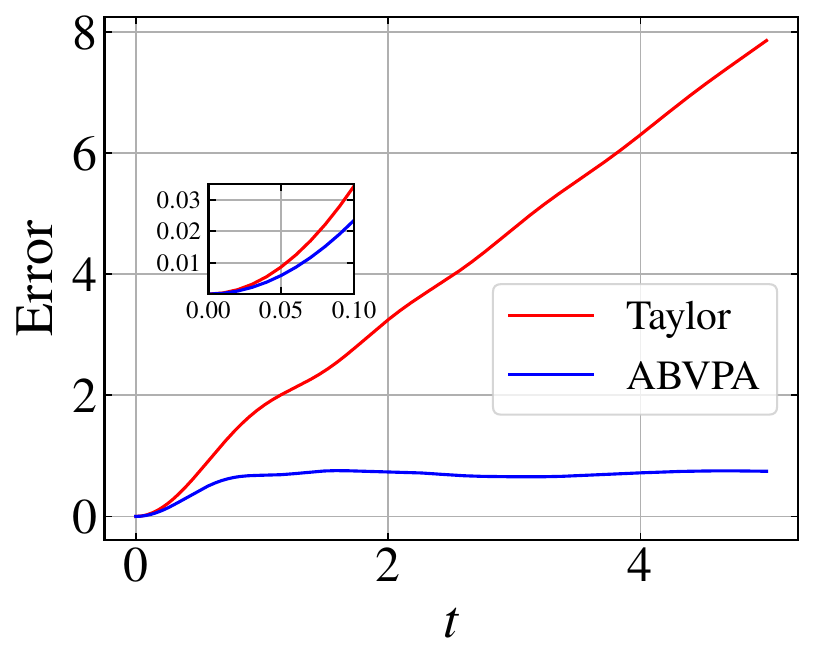}
    \caption{Comparison of the normalized Frobenius-norm error $\frac{1}{2\sqrt{10}}\|e^{-itH}-U_{\rm approx}\|_F$ as a function of $t$, where $U_{\rm approx}$ denotes either the fixed-coefficient Taylor projector $\mathbb{1}-itH$ or the real-time ABVPA, $c_0(t)\mathbb{1}+c_1(t)H$. The results are averaged over 500 random $10\times 10$ Hamiltonians.}
    \label{fig:error_taylor_vs_var_realTE}
\end{figure}

Our second time-evolution ansatz will be used to obtain an improved version of the second-order TS formula. For a Hamiltonian $H=A+B$ where $A$ and $B$ are Hermitian non-commuting operators, we consider the following time-evolution action-based variational TS ansatz (ABVTSA) \cite{Assi2026}
\begin{equation}
\label{eqn:TS_var_ansatz2}
U_a(t)=e^{id_0(t)A}e^{id_1(t)B}e^{id_2(t)A}.
\end{equation}
The corresponding EOM for $\{d_0(t),d_1(t),d_2(t)\}$ are obtained from Eq. \eqref{eq:real_t_EOM} to be
\begin{align}
\label{eqn:ds_for_var_TS}
    g_{00} \Dot{d_0} + g_{01} \Dot{d_1} + g_{02} \Dot{d_2} = D_0, \\
    g_{10} \Dot{d_0} + g_{11} \Dot{d_1} + g_{12} \Dot{d_2} = D_1, \\g_{20} \Dot{d_0} + g_{21} \Dot{d_1} + g_{22} \Dot{d_2} = D_2,  
\end{align}
where $g_{ij}$ are the components of the quantum geometric tensor $g$
\begin{equation}
    g=\begin{bmatrix}
        \mathrm{Tr}[A^2]&\mathrm{Tr}[AB]& T_1(d_1)\\
        \mathrm{Tr}[AB]&\mathrm{Tr}[B^2]& \mathrm{Tr}[AB]\\
        T_1(d_1)&\mathrm{Tr}[AB]& \mathrm{Tr}[A^2]\\
    \end{bmatrix},
\end{equation}
where
\begin{equation}
    T_1(d_1)=\mathrm{Tr}\left[Ae^{id_1B}Ae^{-id_1B}\right],
\end{equation}
and,
\begin{equation}
\begin{aligned}
D_0 &= -\mathrm{Tr} \left[A^2\right]-\mathrm{Tr} \left[AB\right], \\
D_1 &= -\mathrm{Tr} \left[AB\right]-\mathrm{Tr} \left[Be^{id_0A}Be^{-id_0A}\right], \\
D_2 &= -T_1(d_1)-\mathrm{Tr} \left[e^{id_1B}Ae^{-id_1B}e^{-id_0A}Be^{id_0A}\right],
\end{aligned}
\end{equation}
with
\begin{equation}
    T_2(d_0)=\mathrm{Tr} \left[Be^{id_0A}Be^{-id_0A}\right],
\end{equation}
and
\begin{equation}
    T_3(d_0,d_1)=\mathrm{Tr} \left[e^{-id_0A}Be^{id_0A}e^{id_1B}Ae^{-id_1B}\right].
\end{equation}
The above static traces for the QIM have been obtained analytically in Appendix \ref{app:Traces}. Furthermore, the dynamical traces were also obtained analytically \cite{Assi2026}
\begin{equation}
    T_1(d_1)=h_x^2 \, 2^{N-2} \cos(d_1 h_z) \cos\!\left(\frac{d_1 J}{2}\right) \left[ (N-2) \cos\!\left(\frac{d_1 J}{2}\right) + 2 \right],
\end{equation}

\begin{equation}
    T_2(d_0)=\frac{J^22^{N}(N-1)}{16}\cos^2(h_xd_0)+\frac{h_z^22^{N}N}{4}\cos(h_x d_0),
\end{equation}

and

\begin{align}
T_3(d_0,d_1)&=h_x 2^{N-2} \sin(h_x d_0) \Biggl\{ 
h_z \sin(d_1 h_z) \cos\!\left(\frac{d_1 J}{2}\right) \left[2 + (N-2)\cos\!\left(\frac{d_1 J}{2}\right)\right]\nonumber \\
&\qquad + J \cos(h_x d_0) \cos(d_1 h_z) \sin\!\left(\frac{d_1 J}{2}\right) \left[1 + (N-2)\cos\!\left(\frac{d_1 J}{2}\right)\right] \Biggr\}.
\end{align}
These analytic formulas of the traces surpass the need to directly deal with exponentially large Hilbert spaces and they can be used automatically regardless of the qubit counts or the Hamiltonian parameters.

\section{Error bounds for the TS and ABVTSA}
\label{app:TS_errors}
In this appendix, we derive the leading short-time global errors and residual-based error bounds for the r TS decomposition and the ABVTSA construction. Starting from the QIM in Eq.~\eqref{eqn:QIM} and its decomposition $H=A+B$ in Eq.~\eqref{eqn:HamiltonianParts}, we identify the leading error operators for both approximations and obtain the corresponding characteristic time scales.

\subsection{TS error}
The TS propagator given by \ref{eqn:TS_formula}. Using the Baker--Campbell--Hausdorff expansion, its difference from the exact propagator is
\begin{equation}
\label{eqn:TS_leading_difference}
e^{-it(A+B)}-U_{\rm TS}(t)
=\frac{it^3}{24}C_{\rm TS}
+\mathcal{O}(t^4),
\end{equation}
where
\begin{equation}
\label{eqn:C_TS}
C_{\rm TS} = [A,[A,B]] + 2[B,[A,B]].
\end{equation}
It follows that the leading global error is
\begin{equation}
\label{eqn:TS_global_short}
\epsilon_{\rm TS}(t) = \frac{t^3}{48\sqrt{D}} \left\|C_{\rm TS}\right\|_F + \mathcal{O}(t^4).
\end{equation}
The corresponding residual satisfies
\begin{equation}
\label{eqn:TS_residual_short}
R_{\rm TS}(t)=\frac{t^2}{8}C_{\rm TS}+\mathcal{O}(t^3).
\end{equation}
Therefore,
\begin{equation}
\begin{aligned}
\epsilon_{\rm TS}^*(t)
&=\frac{1}{2\sqrt{D}} \int_0^t ds\, \left\|R_{\rm TS}(s)\right\|_F \\
&=\frac{t^3}{48\sqrt{D}}\left\|C_{\rm TS}\right\|_F+\mathcal{O}(t^4).
\end{aligned}
\end{equation}
Introducing
\begin{equation}
q_{\rm TS} =\frac{1}{D}\left\|C_{\rm TS}\right\|_F^2,
\end{equation}
the leading error bound can be written as
\begin{equation}
\label{eqn:TS_q_error}
\epsilon_{\rm TS}^*(t)=\frac{\sqrt{q_{\rm TS}}}{48}t^3+\mathcal{O}(t^4).
\end{equation}
For the QIM considered here, direct evaluation of the commutators gives
\begin{equation}
\label{eqn:q_TS_QIM}
\begin{aligned}
q_{\rm TS}=h_x^2\Bigg[
& \frac{(4N-7)J_z^4}{32}+2(N-1)J_z^2h_x^2\\
&+\frac{3}{4}(N-1)J_z^2h_z^2+Nh_x^2h_z^2+\frac{N}{4}h_z^4\Bigg].
\end{aligned}
\end{equation}
Defining
\begin{equation}
\epsilon_{\rm TS}^*(t)\simeq\left(\frac{t}{t_{\rm TS}^*}\right)^3,
\end{equation}
we obtain
\begin{equation}
\label{eqn:t_star_TS}
t_{\rm TS}^*=\left(\frac{48}{\sqrt{q_{\rm TS}}}\right)^{1/3}.
\end{equation}
\subsection{ABVTSA error}

The ABVTSA and the short-time expansions of its parameters were introduced in Eqs.~\eqref{eqn:TS_var_ansatz}--\eqref{eqn:VTSC}. 
For the QIM decomposition considered here, the operators $A$ and $B$ contain only $Z$-type and $X$-type Pauli strings, respectively, and therefore
\begin{equation}
    \mathrm{Tr}[AB]=0.
\end{equation}
Substituting Eqs.~\eqref{eqn:var_p1}--\eqref{eqn:var_p4} into Eq.~\eqref{eqn:VTSC}, we obtain
\begin{equation}
\label{eqn:d03_QIM}
    d_0^{(3)}(0)=d_2^{(3)}(0)=-\frac{h_x^2\left[J^2(N-1)+2Nh_z^2\right]}{2\left[J^2(N-1)+4Nh_z^2\right]},
\end{equation}
and
\begin{equation}
\label{eqn:d13_QIM}
    d_1^{(3)}(0)=\frac{J^2(N-1)+2Nh_z^2}{8N}.
\end{equation}
Using the Baker--Campbell--Hausdorff expansion together with Eq.~\eqref{eqn:approx_cs_3exp}, the leading error operator for the ABVTSA ansatz is
\begin{equation}
\label{eqn:C_VTS}
    C_{\rm ABVTSA}=C_{\rm TS}-8d_0^{(3)}(0)A-4d_1^{(3)}(0)B,
\end{equation}
where $C_{\rm TS}$ is the leading error operator of the  TS decomposition defined in the preceding subsection. Substituting Eqs.~\eqref{eqn:d03_QIM} and \eqref{eqn:d13_QIM} gives
\begin{equation}
\label{eqn:C_VTS_QIM}
    C_{\rm ABVTSA}=C_{\rm TS}+\frac{4h_x^2\left[J^2(N-1)+2Nh_z^2\right]}{J^2(N-1)+4Nh_z^2}A-\frac{J^2(N-1)+2Nh_z^2}{2N}B.
\end{equation}

The difference between the exact propagator and the ABVTSA approximation is
\begin{equation}
\label{eqn:VTS_leading_difference}
    e^{-itH}-U_{\rm ABVTSA}(t)=\frac{it^3}{24}C_{\rm ABVTSA}+\mathcal{O}(t^4).
\end{equation}
The corresponding leading global error is therefore
\begin{equation}
\label{eqn:VTS_global_error}
    \epsilon_{\rm ABVTSA}(t)=\frac{t^3}{48\sqrt{D}}\left\|C_{\rm ABVTSA}\right\|_F+\mathcal{O}(t^4).
\end{equation}
Similarly, the ABVTSA residual satisfies
\begin{equation}
\label{eqn:VTS_residual}
    R_{\rm ABVTSA}(t)=\frac{t^2}{8}C_{\rm ABVTSA}+\mathcal{O}(t^3),
\end{equation}
and the residual-based error bound becomes
\begin{equation}
\label{eqn:VTS_error_bound}
    \epsilon_{\rm ABVTSA}^*(t)=\frac{t^3}{48\sqrt{D}}\left\|C_{\rm ABVTSA}\right\|_F+\mathcal{O}(t^4),
\end{equation}
defining
\begin{equation}
    q_{\rm ABVTSA}=\frac{1}{D}\left\|C_{\rm ABVTSA}\right\|_F^2,
\end{equation}
we obtain
\begin{equation}
\label{eqn:q_VTS_QIM}
\begin{aligned}
q_{\rm ABVTSA}=q_{\rm TS}
&-\frac{h_x^4\left[J^2(N-1)+2Nh_z^2\right]^2}{J^2(N-1)+4Nh_z^2}\\
&-\frac{h_x^2\left[J^2(N-1)+2Nh_z^2\right]^2}{16N}.
\end{aligned}
\end{equation}
Thus,
\begin{equation}
    \epsilon_{\rm ABVTSA}^*(t)=\frac{\sqrt{q_{\rm ABVTSA}}}{48}t^3+\mathcal{O}(t^4).
\end{equation}
Since the ABVTSA is a second-order approximation, its leading error scales as $t^3$. We therefore define its characteristic time scale through
\begin{equation}
    \epsilon_{\rm ABVTSA}^*(t)\simeq\left(\frac{t}{t_{\rm ABVTSA}^*}\right)^3,
\end{equation}
which gives
\begin{equation}
\label{eqn:t_star_VTS}
    t_{\rm ABVTSA}^*=\left(\frac{48}{\sqrt{q_{\rm ABVTSA}}}\right)^{1/3}.
\end{equation}
\end{widetext}

\bibliographystyle{unsrt}
\bibliography{ref}
\end{document}